\documentclass[twocolumn,prd,superscriptaddress,preprintnumbers,nofootinbib]{revtex4-2}

\usepackage{comment} 

\usepackage{graphicx}
\usepackage{amsmath,bm,amssymb,amsfonts,dsfont}
\usepackage[usenames,dvipsnames]{xcolor}
\usepackage[normalem]{ulem}
\usepackage{url}
\usepackage{array}
\usepackage{booktabs}
\usepackage{multirow}
\usepackage{float}
\usepackage[colorlinks  = true,
            linkcolor   = NavyBlue,
            urlcolor    = NavyBlue,
            citecolor   = NavyBlue,
            anchorcolor = NavyBlue]{hyperref}
\usepackage{cprotect}

\usepackage{appendix}

\usepackage[switch]{lineno}

\usepackage{tikz-feynman}
\tikzfeynmanset{compat=1.1.0}

\pdfoutput=1 

\newcommand{\lsim}{\mathrel{\mathop{\kern 0pt \rlap
  {\raise.2ex\hbox{$<$}}}
  \lower.9ex\hbox{\kern-.190em $\sim$}}}
\newcommand{\gsim}{\mathrel{\mathop{\kern 0pt \rlap
  {\raise.2ex\hbox{$>$}}}
  \lower.9ex\hbox{\kern-.190em $\sim$}}}

\def \aj  {AJ}

\def \jcap  {JCAP}

\def \mnras {MNRAS}

\tikzfeynmanset{warn luatex=false} 

\begin{document}



\title{Strong Constraints on Millisecond Pulsar Injection Spectra \\ from Fermi-LAT Observations of the Galactic Center}

\author{Jordan Koechler}
\email{jordan.koechler@gmail.com}
\affiliation{Istituto Nazionale di Fisica Nucleare, Sezione di Torino, Via P. Giuria 1, 10125 Torino, Italy}

\author{Pedro De la Torre Luque}
\email{pedro.delatorre@uam.es}
\affiliation{Departamento de F\'isica Te\'orica, M-15, Universidad Aut\'onoma de Madrid, E-28049 Madrid, Spain}
\affiliation{Instituto de F\'isica Te\'orica UAM-CSIC, Universidad Aut\'onoma de Madrid, C/ Nicol\'as Cabrera, 13-15, 28049 Madrid, Spain}

\author{Mattia Di Mauro}
\email{dimauro.mattia@gmail.com}
\affiliation{Istituto Nazionale di Fisica Nucleare, Sezione di Torino, Via P. Giuria 1, 10125 Torino, Italy}

\date{\today}

\begin{abstract}

Millisecond pulsars (MSPs) are a leading explanation of the Galactic Center excess (GCE) observed in \textsc{Fermi}-LAT data. We constrain this scenario by jointly modeling prompt and inverse-Compton $\gamma$ rays from MSP-injected $e^\pm$ on the Galactic bulge, using recent \textsc{Fermi}-LAT GCE spectra from state-of-the-art interstellar emission models and data analysis. Current data place strong upper limits on the efficiency ratio $\eta_e/\eta_\gamma$ across broad $e^\pm$ injection scenarios, surpassing those from globular-cluster observations with \textsc{Magic} and competitive with projected \textsc{Ctao} sensitivities toward the Galactic bulge.

\end{abstract}


\maketitle

\flushbottom

\emph{Introduction—} The Galactic Center (GC), expected to contain the largest dark matter (DM) density in the local Universe~\cite{Pieri:2009je}, is thus a particularly compelling target for indirect detection.
A number of analyses of \textsc{Fermi}-LAT data have found an excess of GeV $\gamma$-rays relative to conventional diffuse-emission expectations, known as the Galactic Center Excess (GCE)~\cite{Goodenough:2009gk,Hooper:2010mq,Boyarsky:2010dr,Hooper:2011ti,Abazajian:2012pn,Gordon:2013vta,Abazajian:2014fta,Daylan:2014rsa,Calore:2014nla,Calore:2014xka,TheFermi-LAT:2015kwa,Linden:2016rcf,TheFermi-LAT:2017vmf,DiMauro:2019frs,DiMauro:2021raz,Cholis:2021rpp}.
The GCE exhibits a spectrum peaking at a few GeV and extending up to $\sim\!50$~GeV, together with a spatial morphology well fitted by a generalized NFW profile with inner slope $\gamma \simeq 1.2$ (see, e.g.,~\cite{Calore:2014nla,DiMauro:2021raz,Cholis:2021rpp}).
Such properties are compatible with a DM interpretation in terms of annihilating WIMPs, for example particles with mass $m_{\rm DM}\sim 30$--$60$~GeV annihilating into hadronic~\cite{Calore:2014xka,DiMauro:2021qcf,DiMauro:2023tho,Kong:2025ccv} or leptonic~\cite{Koechler:2025ryv,Kong:2025ccv} final states, with an annihilation cross section close to the canonical thermal value.

\begin{figure}[t!]
    \centering
    \includegraphics[width=\linewidth]{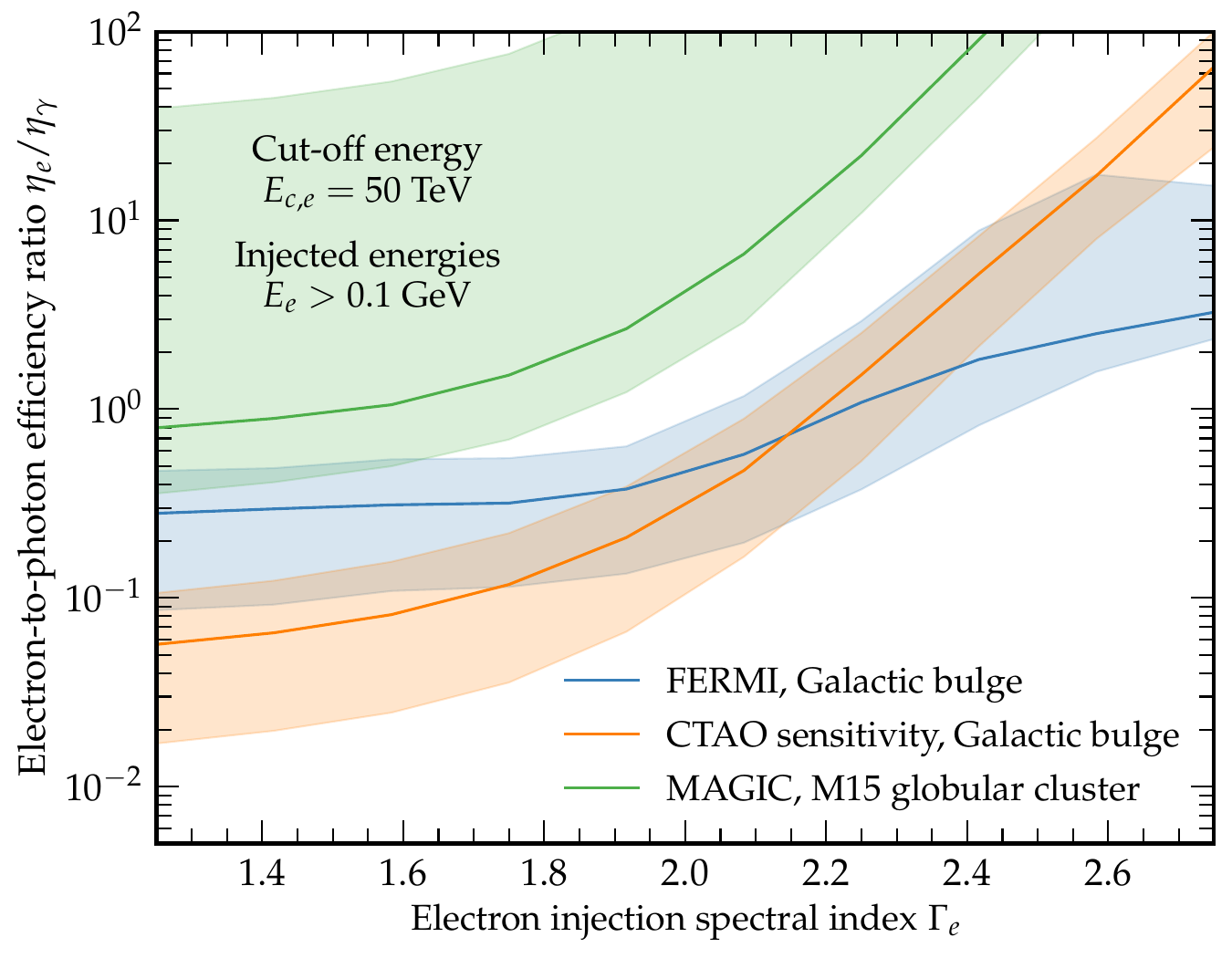}
    \caption{\textbf{Upper limits (95\% C.L.) on the ratio of electron-to-photon efficiencies $\eta_e/\eta_\gamma$, for MSPs under different electron injection scenarios}. The solid blue line shows the constraints derived from \textsc{Fermi}-LAT observations using the Pohl22 IEM, while the orange line includes the projected sensitivity of \textsc{Ctao}. The green line, shown for reference, corresponds to the best current upper limits, derived from \textsc{Magic} observations of the M15 globular cluster~\cite{MAGIC:2019aof}. The colored bands represent the corresponding systematic uncertainties, as explained in the text.} 
    \label{fig:moneyplot}
\end{figure}

Millisecond pulsars (MSPs) have long been proposed as compelling astrophysical sources that could explain the GCE~\cite{Bartels:2015aea,Lee:2015fea,Macias:2016nev,Bartels:2017vsx,Manconi:2024tgh}. Beyond their prompt $\gamma$-ray emission, MSPs can be efficient accelerators of electron-positron pairs ($e^\pm$), which generate very-high-energy $\gamma$ rays through inverse-Compton scattering (ICS) on the interstellar radiation field (ISRF). This makes $\gamma$-ray observations towards the GC a direct probe of the $e^\pm$ injection from MSPs, thereby providing important constraints on their emission properties that have not been fully explored so far~\cite{Macias:2021boz,Keith:2022xbd}.

The $\gamma$-ray emission from the inner Galaxy is overwhelmingly dominated by foreground Galactic interstellar emission, which accounts for approximately 90\% of the observed flux, while the GCE contributes only 5--10\%. Consequently, any attempt to characterize the spectrum or morphology of the GCE is intrinsically limited by the accuracy of the interstellar emission models (IEMs) employed~\cite{Leane:2019uhc,Chang:2019ars,Zhong:2019ycb,Calore:2021jvg,List:2025qbx}. This limitation is particularly relevant at high energies, where the choice of IEM strongly affects the inferred high-energy tail of the GCE, which is crucial for studying possible contributions from MSPs.

Over the past years, the development of more sophisticated \textsc{Fermi}-LAT analysis methods has substantially reduced the impact of systematic uncertainties associated with the choice of IEM. Building on this progress, a recent analysis in Ref.~\cite{DiMauro:2026fnp} examined a suite of physically motivated IEMs and employed an improved treatment of the sources in the GC region, enabling a more accurate extraction of the GCE properties. In this reference the residuals between model and data have been found to be at most about $10\%$, i.e.~significantly smaller than previous papers (see, e.g., \cite{Cholis:2021rpp,DiMauro:2021raz,Pohl:2022nnd}). Among the most relevant results for the MSP interpretation are robust upper limits at the level of $E^2\Phi \lesssim 10^{-8}$~GeV\,cm$^{-2}$\,s$^{-1}$\,sr$^{-1}$ for $\gamma$-ray energies of $E>10$~GeV, where the ICS component can dominate over the prompt emission. Based on this procedure, our analysis provides the most robust limits obtained so far on the ratio between the efficiency for converting MSP spin-down luminosity into $e^\pm$ and that into prompt $\gamma$-rays ($\eta_e/\eta_\gamma$), which are shown as the blue band in Fig.~\ref{fig:moneyplot}.

Previous constraints on $e^\pm$ injection from MSPs have been limited and have primarily focused on globular clusters rather than the GC.
Observations with Cherenkov telescopes such as \textsc{Hess}, \textsc{Magic}, and \textsc{Veritas} have targeted multiple globular clusters~\cite{Abramowski2011Terzan5, Abramowski2013GCs, Anderhub2009M13, McCutcheon2009VERITAS, MAGIC:2019aof}, but have mostly resulted in non-detections, thereby providing upper limits on the $\gamma$-ray flux and constraints on particle injection. Deep observations of the globular cluster M15 with \textsc{Magic}~\cite{MAGIC:2019aof} have yielded the strongest current constraints on the efficiency with which MSPs convert rotational energy into relativistic $e^\pm$. However, these constraints are affected by significant uncertainties, since the relevant environmental conditions (magnetic fields, winds, and particle propagation) are poorly known~\cite{Ndiyavala-Davids:2020wjc}. We revisited these upper limits on $\eta_e/\eta_\gamma$, using the same method used for the \textsc{Fermi}-LAT non-detection upper limits at the GC, and obtained the green band in Fig.~\ref{fig:moneyplot}. 
In particular, our upper limits on $\eta_e/\eta_\gamma$ are lower by a factor of 4-100 with respect to current bounds from \textsc{Magic} \cite{MAGIC:2019aof}, which  represents a significant advancement towards testing the MSP interpretation of the GCE.


For young rotation-powered pulsars, the apparent \textit{Fermi}-LAT $\gamma$-ray efficiency spans a broad range. However, after accounting for beaming corrections, a representative intrinsic prompt efficiency is $\eta_\gamma \sim 0.02$--$0.08$. Combined with lepton efficiencies of $\eta_e \sim 0.05$--$0.10$ for the escaping $e^\pm$ component relevant for ICS emission, or $\eta_e \sim 0.10$--$0.30$ for the total lepton output of young pulsars and their associated pulsar wind nebulae, this implies $\eta_e/\eta_\gamma \sim 1$--$5$ and $\sim 2$--$10$, respectively~\cite{Fermi-LAT:2023zzt,Pierbattista:2014ona,Albert:2024,Olmi:2023zal} (see Appendix~\ref{appx:youngpulsars} for further details).


Meanwhile, the future Cherenkov Telescope Array Observatory (\textsc{Ctao}), with its unprecedented sensitivity, angular resolution, and TeV energy coverage, will offer significant opportunities to detect or constrain high-energy emission from MSP populations in the GC. While most \textsc{Ctao} sensitivity studies of the GC have focused on DM annihilation~\cite{CTA:2020qlo}, the high-resolution survey of the inner Galaxy also enhances the prospects for studying diffuse or unresolved components, including ICS emission from MSPs. Forecasts suggest that \textsc{Ctao} will place strong limits on particle injection efficiencies and emission models~\cite{Macias:2021boz,Keith:2022xbd}. 
Remarkably, we show in Fig.~\ref{fig:moneyplot} that current \textsc{Fermi}-LAT observations of the GC assuming $e^\pm$ injection spectral index $\Gamma_e\gtrsim1.9$ are comparable or stronger than \textsc{Ctao} sensitivity prospects. This result stands for all physically motivated $e^\pm$ injection cut-off energies. For harder spectra, which are less physically motivated for MSPs~\cite{Wang:2006mspgc,Venter:2014ata,Petrovic:2014xra,MAGIC:2019aof,Ndiyavala-Davids:2020wjc,Macias:2021boz}, the constraints obtained with \textsc{Fermi}-LAT can be at most a factor of 2 weaker than \textsc{Ctao}. This demonstrates the unique power of existing \textsc{Fermi}-LAT $\gamma$-ray data to probe MSP activity in the inner Galaxy.

This Letter is structured as follows: (i) we describe the theoretical framework used to predict the photon flux from the GC and M15; (ii) we present the datasets employed in this study; (iii) we outline the analysis pipeline; (iv) we present the results; and (v) we conclude.

\emph{Theoretical setup—}
We model the injection spectrum of prompt $\gamma$-rays and $e^\pm$ from MSPs as a power law with an exponential cut-off,
$d^2N/(dE\,dt)\propto E^{-\Gamma}\exp(-E/E_c)$.
To compute the total photon flux from MSPs in the Galactic bulge (GB), we take into account both prompt emission and secondary emission arising from the up-scattering of ambient photons by the $e^\pm$ produced by MSPs, i.e.~ICS. The prompt photon flux is given by (see the full derivation in Appx.~\ref{appx:setup})
\begin{equation}
    \label{eq:promptsimp}
    \frac{d\Phi_\gamma^{\rm prompt}}{dE_\gamma} =\frac{\mathcal{D}}{4\pi}\frac{L_{\gamma,\rm tot}}{L_\gamma}\,\frac{d^2N_\gamma}{dE_\gamma dt}\;,
\end{equation}
where $\mathcal{D} = \int_{\Delta\Omega}d\Omega\int_{\rm l.o.s.}ds\,(\rho_{\rm bulge}(r(s,\Omega))/M_{\rm bulge})$ is analogous to the $D$-factor used in DM studies. Here, $L_{\gamma,\rm tot}$ is the total intrinsic luminosity of the GB in prompt photons, while
$L_\gamma=\int dE_\gamma\,E_\gamma\,d^2N_\gamma/(dE_\gamma dt)$ is the prompt photon luminosity of an individual MSP. Finally, $\rho_{\rm bulge}$ is the GB mass density and
$M_{\rm bulge}=\int \rho_{\rm bulge}\,dV$ is the total mass of the GB.

The ICS photon flux can be written as (see the full derivation in Appx.~\ref{appx:setup})
\begin{equation}
    \label{eq:ICSsimp}
    \frac{d\Phi_\gamma^{\rm ICS}}{dE_\gamma} = \frac{\mathcal{D}}{4\pi E_\gamma}\frac{\eta_e}{\eta_\gamma}\frac{L_{\gamma,\rm tot}}{L_e}\,\int_{E_e^{\rm min}}^\infty dE_e\frac{\mathcal{P}_{\rm ICS}(E_\gamma,E_e)}{b(E_e)}\mathcal{Y}(E_e)\,,
\end{equation}
where $\mathcal{Y}(E_e) = \int_{E_e}^\infty dE_e' \, d^2N_e/(dE_e' dt)$,
$L_e=\int dE_e\,E_e\,d^2N_e/(dE_e dt)$ is the $e^\pm$ luminosity of an individual MSP, $E_e^{\rm min}=0.1$ GeV, $\mathcal{P}_{\rm ICS}(E_\gamma,E_e)$ is the differential ICS power emitted by a single $e^\pm$~\cite{Cirelli:2009vg}, and
$b(E_e)\equiv -dE_e/dt$ is the total $e^\pm$ energy-loss rate in the Galactic medium.
To obtain this simplified expression we assume that, in the inner Galaxy, energy losses (ICS + synchrotron) dominate over spatial diffusion, as done in Refs.~\cite{DiMauro:2015tfa,Cirelli:2009vg,Blanchet:2012vq,DiMauro:2023oqx}, and that the Galactic magnetic field (GMF) and ISRFs (hence $b$ and $\mathcal{P}_{\rm ICS}$) are spatially homogeneous. We adopt a constant GMF of $5.4~\mu$G, corresponding to the average field strength within the $40^\circ\times40^\circ$ ROI, following the GMF models of Refs.~\cite{Evoli:2016xgn,Strong:1998fr}. However, since they are highly uncertain, we allow values between 3 and 7 $\mu$G, corresponding to the lower and upper bounds of the average field strengths predicted by the Galactic magnetic field models considered in Ref.~\cite{Buch:2015iya}. In this setup, for $e^\pm$ energies $E_e \lesssim 200$~GeV, ICS losses dominate over synchrotron losses. At higher energies, ICS enters the Klein--Nishina regime, and the corresponding energy losses become subdominant with respect to synchrotron losses.

In addition to the contribution of the cosmic microwave background (CMB), we model the other ISRF fields using the templates provided in the cosmic-ray propagation code \texttt{GALPROP}~\cite{Porter:2017vaa}. These templates are based on two distinct models: F98~\cite{Freudenreich:1997bx}, which describes emission from the Galactic bar as well as the stellar and dust disks, calibrated to infrared observations from \textsc{Cobe}/\textsc{Dirbe}, and R12~\cite{Robitaille:2012kg}, which includes additional structural components such as the bulge, halo, and spiral arms. Overall, the two models yield similar photon field densities, with differences typically within factors of $\sim 0.1$--$3$, depending on the photon energy.

We emphasize that the total prompt photon luminosity in the GB, $L_{\gamma,\rm tot}$ and the efficiency ratio $\eta_e/\eta_\gamma$ are partially degenerate. To alleviate this degeneracy, we impose the additional requirement that the predicted total photon flux reproduces the luminosity of the GCE measured by \textsc{Fermi}-LAT $L_{\rm GCE}^{\rm meas}$:
\begin{equation}
\label{eq:LGCE}
L_{\rm GCE}^{\rm meas} = 4\pi r_\odot^2 \int_{\Delta E_{\rm GCE}} dE_\gamma\, E_\gamma \left(\frac{d\Phi^{\rm prompt}_\gamma}{dE_\gamma}+\frac{d\Phi^{\rm ICS}_\gamma}{dE_\gamma}\right)\;,
\end{equation}
where $r_\odot$ is the Galactocentric distance. In the approximation in which diffusion effects are neglected, Eqs.~(\ref{eq:promptsimp}--\ref{eq:LGCE}) give $L_{\gamma,\rm tot}\propto L_{\rm GCE}^{\rm meas}/\mathcal{D}$. 
Expressing $L_{\gamma,\rm tot}$ in terms of $L_{\rm GCE}^{\rm meas}$ therefore removes the explicit dependence on $\mathcal{D}$ in the predicted flux. This, in turn, allows us to avoid specifying a particular spatial model for the GB density $\rho_{\rm bulge}$ in the theoretical prediction. That dependence is, however, encoded in the extraction pipeline of the \textsc{Fermi}-LAT data, and therefore implicitly in the value of $L_{\rm GCE}^{\rm meas}$, as explained in the following paragraph. We refer the reader to Eqs.~\ref{eq:fluxtotsimp1} and \ref{eq:fluxtotsimp2} where we write the simplified expressions of the total photon flux emitted by the GB as a function of $L_{\rm GCE}^{\rm meas}$. 

\emph{GCE Dataset—}
Accurate IEMs and a dedicated data analysis are critical to test whether the GCE traces stellar structures such as the nuclear bulge or boxy bulge, a key signature of an MSP origin, and to determine the excess spectrum accurately. Given the dominant role of the IEM, systematic uncertainties directly propagate into the inferred GCE properties. To mitigate this effect, we make use of the results of Ref.~\cite{DiMauro:2026fnp}, where the modeling of \textsc{Fermi}-LAT data was significantly improved with respect to previous analyses. In particular, we select the IEMs that provide the best agreement with the data: Macias18~\cite{Macias:2016nev}, Galp21 (SA0, SA50, SA100)~\cite{Porter:2017vaa} and Pohl22~\cite{Pohl:2022nnd} (see more details in Appx.~\ref{appx:IEMs}).

While the Galp21 models yield the lowest residuals in the GCE analysis, the Macias18 and Pohl22 models provide the highest significance for the excess when boxy-bulge templates are used. Based on this, the Pohl22 template is the most appropriate for studies of the MSP contribution, although we test all the models in order to assess the associated systematics (e.g., from the \textsc{Fermi} bubbles templates). For each IEM we can also infer the value of the $L_{\rm GCE}^{\rm meas}$ that is of the order of $(5-7) \times 10^{36}$ erg/s (see Appx.~\ref{appx:IEMs}).

Another aspect of our analysis is to explore the prospects for future GCE measurements with \textsc{Ctao}. To date, no official \textsc{Ctao} publication has presented sensitivity projections specifically for MSPs in the Galactic bulge. One possible approach to address this limitation is to simulate observations based on the expected instrumental performance~\cite{Keith:2022xbd,Macias:2021boz}. Instead, we adopt a data-driven strategy and make use of the publicly available \textsc{Ctao} sensitivity to DM annihilation in a $12^\circ\times12^\circ$ region around the GC~\cite{CTA:2020qlo}, provided in the Zenodo repository~\cite{bringmann_2020_4057987}. From these results, we extract the 95\% confidence-level upper limits on the DM-induced signal and reinterpret them in the context of MSP emission.

\emph{Analysis—}
The parameters governing the total $\gamma$-ray flux from MSPs in the GB are as follows: the prompt $\gamma$-ray and $e^\pm$ injection parameters $(\Gamma_\gamma, E_{c,\gamma}, \Gamma_e, E_{c,e})$, which determine the spectral shape, and the ratio of efficiencies $\eta_e/\eta_\gamma$, which controls the relative normalization of the ICS and prompt components. 

In our analysis, we treat $\Gamma_e$ and $E_{c,e}$ as nuisance parameters. This choice reflects the fact that the injection spectrum of $e^\pm$ from MSPs is only weakly constrained by \textsc{Fermi}-LAT data. In particular, the primary observable sensitive to $e^\pm$ injection is the ICS emission, which dominates over the prompt component at $E_\gamma \gtrsim 20$~GeV. In this energy range, however, \textsc{Fermi}-LAT data are largely limited to upper bounds, resulting in weak constraints on the electron injection parameters.

For each dataset described before and for a given $e^\pm$ injection scenario, we perform a fit of the parameters $\Gamma_\gamma$, $E_{c,\gamma}$, and $\eta_e/\eta_\gamma$. To this end, we use the Monte Carlo Markov Chain (MCMC) package \texttt{emcee}~\cite{Foreman-Mackey:2012any} to sample and maximize the following log-likelihood function, $\log\mathcal{L}$:
\begin{multline}
\label{eq:loglikelihood}
-2\log\mathcal{L}=\sum_{i\in\rm meas}\left(\frac{\Phi_{\gamma,i}^{\rm MSP}-\Phi_{\gamma,i}^{\rm meas}}{\sigma_i}\right)^2+\\
+\sum_{i\in\rm UL}\left(\frac{\max(\Phi_{\gamma,i}^{\rm MSP}-\Phi_{\gamma,i}^{\rm UL},0)}{\sigma_i}\right)^2\;,
\end{multline}
where the first sum runs over energy bins in which \textsc{Fermi}-LAT reports a detection, while the second sum includes bins where only upper limits are provided. We adopt flat priors for all fitted parameters: $-2<\log_{10}(\eta_e/\eta_\gamma)<1$, $0<\Gamma_\gamma<2$, and $0<E_{c,\gamma}/\mathrm{GeV}<5$. Within this MCMC framework, we determine the best-fit values of $\Gamma_\gamma$ and $E_{c,\gamma}$, as well as the 95\% C.L.\ upper limit on $\eta_e/\eta_\gamma$, for each of the considered \textsc{Fermi}-LAT IEMs and $e^\pm$ injection scenarios.

\begin{figure}
    \centering
    \includegraphics[width=\linewidth]{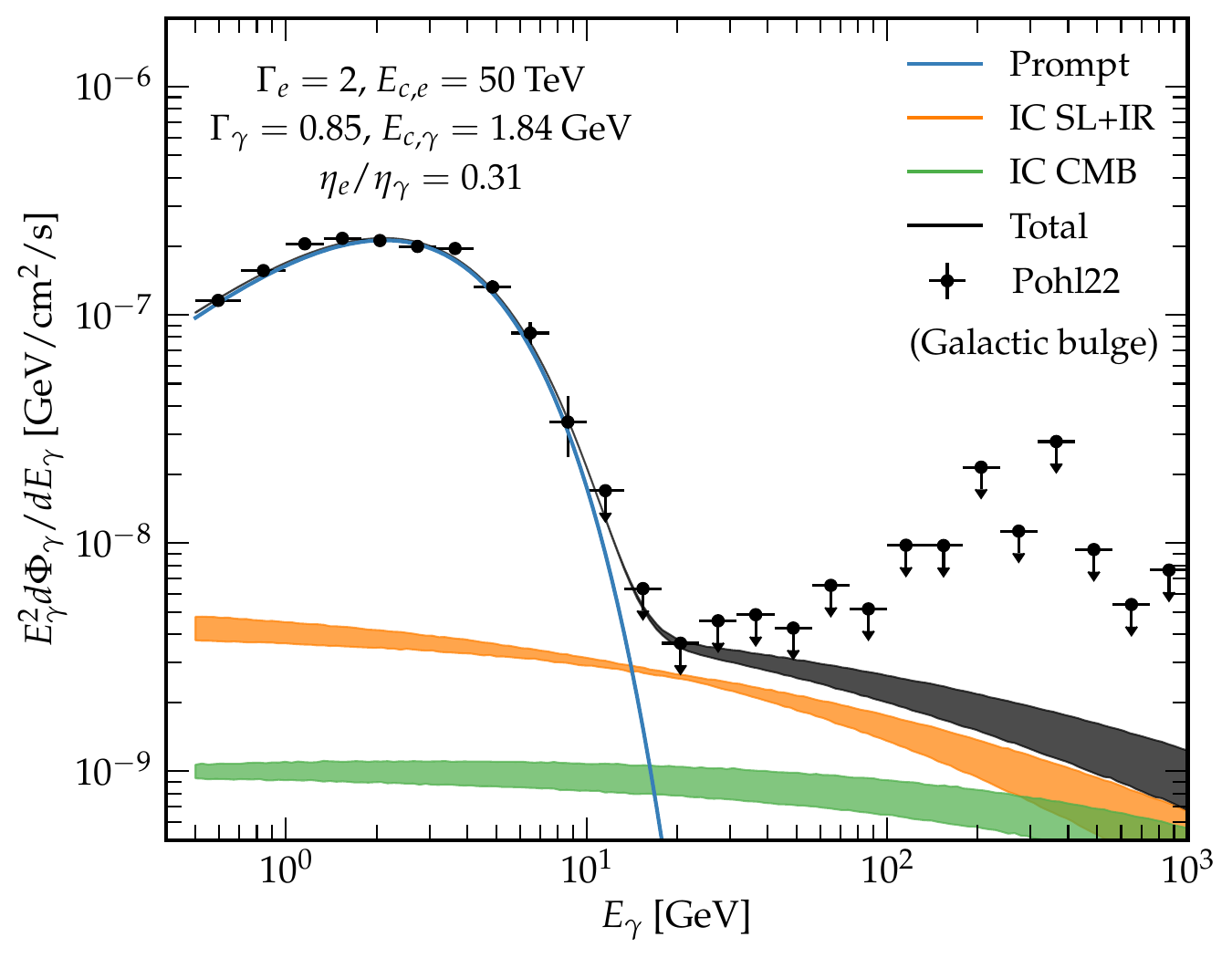}
    \caption{\textbf{Predicted total photon flux (black band) from MSPs in the GB, fitted to the \textsc{Fermi}-LAT data extracted using the Pohl22 IEM (black points).} The solid blue line represents the prompt emission, while the orange and green bands show the systematic uncertainties of the IC fluxes, related to the choice of the starlight (SL) and infrared (IR) photon field model (F98 and R12) and the magnetic field (3 to 7 $\mu$G).}
    \label{fig:FitExample}
\end{figure}

We also derive upper limits on $\eta_e/\eta_\gamma$ for forthcoming observations of the GC with \textsc{Ctao}, following the setup described in Ref.~\cite{CTA:2020qlo}.
Given that the \textsc{Ctao} energy range extends from $30$ to $10^5$~GeV, the predicted photon flux in this regime is dominated by ICS. We therefore approximate the total emission as purely ICS and find the value of $\eta_e/\eta_\gamma$ by requiring the predicted ICS flux not to exceed the projected \textsc{Ctao} sensitivity upper limits (using only the second sum of Eq.~\eqref{eq:loglikelihood}). In this procedure, all other parameters (i.e.\ $\Gamma_\gamma$ and $E_{c,\gamma}$) are fixed to their best-fit values obtained from the \textsc{Fermi}-LAT analysis. Repeating this procedure for each IEM used in the \textsc{Fermi}-LAT analysis allows us to estimate the associated systematic uncertainty on the derived upper limits of $\eta_e/\eta_\gamma$.

A further consideration concerns the mismatch in the regions of interest: the \textsc{Fermi}-LAT datasets are extracted over a $40^\circ\times40^\circ$ region, while the \textsc{Ctao} projected sensitivity corresponds to a smaller $12^\circ\times12^\circ$ region. To account for this difference, we rescale the predicted flux using the measured spatial morphology of the GB emission~\cite{DiMauro:2026fnp}. Specifically, we compute the ratio of the flux observed by \textsc{Fermi}-LAT, integrated over the two regions and find that, on average, the $12^\circ\times12^\circ$ region contains approximately $90\%$ of the total emission within the $40^\circ\times40^\circ$ region. This correction factor is applied when comparing the model predictions to the \textsc{Ctao} sensitivity.

Finally, we compare our Galactic bulge limits with those obtained from the non-detection of M15 by \textsc{Magic}~\cite{MAGIC:2019aof}. We compute the ICS flux from M15 using the point-source limit of Eq.~\eqref{eq:ICSsimp}, with the prompt luminosity normalized to the \textsc{Fermi}-LAT energy flux above $100$ MeV reported in the 4FGL \cite{Ballet:2023qzs}, $G_\gamma=(3.57\pm0.58)\times10^{-12}\,{\rm erg\,cm^{-2}\,s^{-1}}$. The limit on $\eta_e/\eta_\gamma$ is obtained by requiring the predicted ICS flux not to exceed the \textsc{Magic} upper limits. We include the main environmental uncertainties by varying the magnetic field over $1$--$30\,\mu{\rm G}$ and by modeling the optical radiation field of M15 as described in Appx.~\ref{appx:setup}.

\emph{Results—}
In Fig.~\ref{fig:FitExample}, we show the GCE spectrum (using the Pohl22 IEM) compared to the MSPs expected emission, containing the prompt and ICS contributions evaluated for the $e^\pm$ injection spectral index $\Gamma_e=2$ and energy cut-off $E_{c,e}=50$~TeV. As the figure shows, the high-energy flux upper limits imply that the electron-to-photon efficiency ratio must be lower than $\sim 0.31$. We note that for a standard value of $\eta_{\gamma}\sim0.1$ this implies that the electron efficiency is lower than a $3\%$, a lower value than usually assumed, but still compatible with the value required by young pulsars to explain positron data~\cite{Orusa:2024ewq}.
These constraints supersede those expected from idealized observations of the GC by \textsc{Ctao}, as shown in Fig.~\ref{fig:moneyplot}, where we use the expected \textsc{Ctao} sensitivity to the GCE emission from Ref.~\cite{CTA:2020qlo} to evaluate the same constraint. However, still \textsc{Ctao} could improve these constraints for low values of $\Gamma_e$, as long as the assumed $e^\pm$ cut-off energy is larger than $\sim10$~TeV (see Fig.~\ref{fig:FERMI_vs_CTAO} in Appx.~\ref{appx:add_results}).
Meanwhile, the upper limits on $\eta_e/\eta_{\gamma}$ obtained from M15 observations by \textsc{Magic} are weaker by a factor between 4 and 100 than the ones obtained by \textsc{Ctao} and \textsc{Fermi}-LAT for all considered $e^\pm$ injection scenarios, which is notable, given that they represent the best current upper limits to date.

Another remarkable point is that the constraints on the efficiency ratio obtained from our analysis allows for strong constraints even in the case of low $e^\pm$ cut-off energies, where \textsc{Magic} or \textsc{Ctao}, based on higher-energy observations, are not able to set competitive constraints. One can see, from the upper left panel of Fig.~\ref{fig:FERMI_vs_CTAO}, that for cut-off energies below $1$~TeV, the \textsc{Fermi}-LAT GC observations outperform \textsc{Ctao} for all spectral index values.

\emph{Conclusions—}
In this work we have derived constraints on the high-energy $e^\pm$ emission from MSPs in the GB under the assumption that MSPs account for the GCE. Our analysis combines the prompt $\gamma$-ray component associated with the MSP population and the ICS emission generated by the injected $e^\pm$, and confronts these predictions with the most recent determinations of the GCE spectrum obtained with state-of-the-art \textsc{Fermi}-LAT analyses. A central ingredient of our study is the treatment of the diffuse foregrounds that we take from the results of Ref.~\cite{DiMauro:2026fnp}: because the inner-Galaxy emission is dominated by interstellar foregrounds, robust limits on any ICS contribution can only be obtained with a set of physically motivated IEMs and with templates matched to the stellar morphology expected in an MSP scenario. This point is essential, since both the normalization and the spectral shape of the extracted GCE depend sensitively on the adopted diffuse model.

Our main result is that present \textsc{Fermi}-LAT observations of the GC already place very stringent constraints on the ratio between the efficiency for converting MSP spin-down power into relativistic $e^\pm$ and that into prompt $\gamma$ rays, $\eta_e/\eta_\gamma$. Over a broad and astrophysically relevant region of parameter space, these limits are more stringent than those obtained from globular cluster observations and, in many cases, are also stronger than the projected reach of idealized \textsc{Ctao} studies. In particular, \textsc{Fermi}-LAT is especially competitive for soft or moderately hard injection spectra and for cutoff energies below the multi-TeV range, where the ICS emission still falls largely within the GeV band and can therefore be efficiently constrained by existing data. \textsc{Ctao} may improve upon these bounds only in the regime of very hard spectra and sufficiently large cutoff energies, where a larger fraction of the emitted power is shifted to TeV energies. In addition, these bounds from \textsc{Fermi}-LAT reach values below the typical ratios $\eta_e/\eta_\gamma\sim1$--$10$ inferred for young pulsars~\cite{Grenier:2015,Olmi:2023zal,Amato:2020,Torres:2014,Albert:2024}.


A second important outcome of this work is that modeling MSP emissions from the GC is more robust than in globular clusters. Constraints derived from systems such as M15 are valuable, but they are affected by large astrophysical uncertainties related to the local magnetic field, transport conditions, radiation environment, and the poorly known MSP population in the cluster. By contrast, the GC analysis directly probes the very population hypothesized to explain the GCE, and does so in the same environment in which the excess is observed. For this reason, the bounds derived here on $\eta_e/\eta_\gamma$ should be regarded as the most relevant current limits on the lepton emission of a bulge MSP population.


We have also shown that, even after accounting for the uncertainties associated with the choice of IEM and interstellar radiation field, the allowed values of $\eta_e/\eta_\gamma$ remain compatible with realistic MSP scenarios. The resulting picture is therefore not one in which the ICS component rules out an MSP origin, but rather one in which it provides a powerful consistency test: the same MSP population invoked to explain the prompt GeV excess cannot inject an arbitrarily large amount of high-energy $e^\pm$ without overproducing the observed inner-Galaxy emission. In this sense, the ICS component adds non-trivial information beyond the prompt fit alone and should be included in any realistic assessment of bulge MSP models.



\emph{Acknowledgments—}
We are grateful to Dan Hooper and Tim Linden for their helpful feedback on the draft. P.D.L.~has been supported by the Juan de la Cierva JDC2022-048916-I grant, funded by MCIU/AEI/10.13039/501100011033 European Union "NextGenerationEU"/PRTR, and is currently supported by Ramón y Cajal RYC2024-048445-I grant, which is funded by MCIU/AEI/10.13039/501100011033 and FSE+. The work of P.D.L.~is also supported by the grants PID2021-125331NB-I00 and CEX2020-001007-S, both funded by MCIN/AEI/10.13039/501100011033 and by “ERDF A way of making Europe”. P.D.L.~also acknowledges the MultiDark Network, ref. RED2022-134411-T. M.D.M.~and J.K.~acknowledge support from the research grant {\sl TAsP (Theoretical Astroparticle Physics)} funded by Istituto Nazionale di Fisica Nucleare (INFN), and from the Italian Ministry of University and Research (MUR), PRIN 2022 ``EXSKALIBUR – Euclid-Cross-SKA: Likelihood Inference Building for Universe’s Research'', Grant No. 20222BBYB9, CUP I53D23000610 0006, and from the European Union -- Next Generation EU. J.K.~acknowledges support from the Italian Space Agency through the ASI INFN agreement n. 2018-28-HH.0: “Partecipazione italiana al GAPS - General AntiParticle Spectrometer”.

\bibliographystyle{apsrev4-1}
\bibliography{paper.bib}

\newpage

\onecolumngrid
\appendix

\section{Estimate of $\eta_e/\eta_\gamma$ for young pulsars.}
\label{appx:youngpulsars}

Here we derive a semi-empirical estimate of the ratio between the efficiency for converting
spin-down power into relativistic leptons, $\eta_e$, and into prompt $\gamma$-rays, $\eta_\gamma$,
for young rotation-powered pulsars.

We define
\begin{equation}
\eta_\gamma \equiv \frac{L_\gamma}{\dot E},
\qquad
L_\gamma = 4\pi f_\Omega d^2 G_\gamma,
\end{equation}
where $\dot E$ is the spin-down luminosity, $G_\gamma$ is the observed phase-averaged
energy flux, $d$ is the distance, and $f_\Omega$ is the beaming correction factor.
The quantity usually plotted in the \textsc{Fermi}-LAT catalogs is therefore the
apparent efficiency
\begin{equation}
\eta_{\gamma,\rm app} \equiv \frac{4\pi d^2 G_\gamma}{\dot E},
\end{equation}
while the intrinsic prompt efficiency is
\begin{equation}
\eta_\gamma = f_\Omega \,\eta_{\gamma,\rm app}.
\end{equation}

For young \textsc{Fermi}-LAT pulsars, the catalog data are well described by the heuristic
scaling
\begin{equation}
L_\gamma \propto \dot E^{1/2},
\end{equation}
more specifically
\begin{equation}
L_\gamma \simeq \left(10^{33}\,{\rm erg\,s^{-1}}\times \dot E\right)^{1/2},
\label{eq:LgdotE}
\end{equation}
with substantial scatter~\cite{Fermi-LAT:2023zzt,Romani:2011,Perera:2013,Pierbattista:2014ona}.
Hence
\begin{equation}
\eta_{\gamma,\rm app}
= \frac{L_\gamma}{\dot E}
\simeq \left(\frac{10^{33}\,{\rm erg\,s^{-1}}}{\dot E}\right)^{1/2}.
\label{eq:etag_app}
\end{equation}

A key correction is the beaming factor.
Light-curve fits of 76 young \textsc{Fermi}-LAT pulsars show that the derived
$\gamma$-ray beaming factors are generally $<1$ and that using these beaming factors
removes the apparently unphysical cases with $L_\gamma>\dot E$~\cite{Pierbattista:2014ona}.
Population studies of non-recycled \textsc{Fermi}-LAT pulsars likewise favor broad beams
with an average solid angle $\Omega_\gamma \sim 3.7\pi$, corresponding to
\begin{equation}
f_\Omega \sim \frac{\Omega_\gamma}{4\pi}\sim 0.9,
\end{equation}
although object-to-object scatter is expected~\cite{Perera:2013}.
To remain conservative, we therefore adopt
\begin{equation}
0.5 \lesssim f_\Omega \lesssim 1.
\label{eq:fo_range}
\end{equation}

Combining Eqs.~(\ref{eq:etag_app}) and (\ref{eq:fo_range}) gives
\begin{equation}
\eta_\gamma \simeq
f_\Omega \left(\frac{10^{33}}{\dot E}\right)^{1/2}.
\end{equation}
For representative young-pulsar spin-down power range between $\dot{E}\in[10^{35},10^{37}]$ erg/s we find a representative range
\begin{equation}
\eta_\gamma^{\rm young} \sim (1-5)\times10^{-2}.
\label{eq:etag_final}
\end{equation}

Estimating $\eta_e$ is less straightforward, because one must distinguish between:
(i) the escaping $e^\pm$ component relevant for ICS halos, and
(ii) the total relativistic lepton channel injected into the young pulsar wind nebula (PWN).

For the escaping/ICS-relevant component, the most direct observational anchors are:
\begin{itemize}
\item the Geminga GeV-halo analysis with \textsc{Fermi}-LAT + \textsc{Hawc}, which finds that
an efficiency of about
\begin{equation}
\eta_e^{\rm esc} \sim 0.01
\end{equation}
is sufficient to reproduce the data~\cite{DiMauro:2019yvh};
\item the recent \textsc{Hawc} analysis, which finds \cite{Albert:2024}
\begin{equation}
\eta_e^{\rm esc} \simeq 0.066 \;\;{\rm (Geminga)},
\qquad
\eta_e^{\rm esc} \simeq 0.051 \;\;{\rm (Monogem)}.
\end{equation}
\end{itemize}
These sources are middle-aged rather than very young, but they provide the cleanest direct
measurements of the lepton channel relevant for extended ICS emission. They suggest a
conservative escaped-pair benchmark
\begin{equation}
\eta_e^{\rm esc}\sim 0.01\text{--}0.07.
\label{eq:etae_esc}
\end{equation}

For the full relativistic lepton channel in young PWMe, the literature points to systematically
larger values. Reviews emphasize that PWNe are very efficient particle accelerators and that
young systems are particle-dominated; in particular, the pulsar can inject into particles
``up to a few tens of percent'' of its spin-down power~\cite{Olmi:2023zal,Amato:2021Crab}.
Likewise, modeling of TeV-detected young PWNe concludes that they are particle-dominated
and far from magnetic equipartition~\cite{Torres:2014}.
This motivates
\begin{equation}
\eta_e^{\rm PWN}\sim 0.05\text{--}0.3
\label{eq:etae_pwn}
\end{equation}
as a reasonable young-pulsar/PWN range for the radiatively relevant lepton channel.

Using Eq.~(\ref{eq:etag_final}), the ratio $\eta_e/\eta_\gamma$ can then be estimated in
two ways.

\paragraph{(a) Escaping ICS-relevant electrons/positrons.}
Combining Eqs.~(\ref{eq:etag_final}) and (\ref{eq:etae_esc}) gives
\begin{equation}
\frac{\eta_e}{\eta_\gamma}
\sim
\frac{0.01\text{--}0.07}{0.01\text{--}0.05}
\;\Rightarrow\;
\frac{\eta_e}{\eta_\gamma}\sim 0.2\text{--}7.
\end{equation}

\paragraph{(b) Total young-PWN relativistic lepton channel.}
Combining Eqs.~(\ref{eq:etag_final}) and (\ref{eq:etae_pwn}) gives
\begin{equation}
\frac{\eta_e}{\eta_\gamma}
\sim
\frac{0.05\text{--}0.3}{0.01\text{--}0.05}
\;\Rightarrow\;
\frac{\eta_e}{\eta_\gamma}\sim 1\text{--}30.
\end{equation}

As a compromise between these two definitions, a practical fiducial range for young pulsars is
\begin{equation}
\boxed{
\frac{\eta_e}{\eta_\gamma}\sim 1\text{--}10
}
\end{equation}

\paragraph{Interpretation.}
The lower part of the range applies when $\eta_e$ is identified with the escaped $e^\pm$
component that powers ICS halos.
The upper part applies when $\eta_e$ is understood as the full relativistic lepton budget
of a young, particle-dominated PWN.
Therefore, for comparisons with extended ICS emission, the most relevant estimate is
\begin{equation}
\frac{\eta_e}{\eta_\gamma}\sim {\rm few},
\end{equation}
whereas for the total young-PWN lepton channel one may reasonably allow values up to
\(\mathcal{O}(10)\).

\section{$\gamma$-ray flux prediction}
\label{appx:setup}

In this appendix, we detail the computation of the photon flux from the MSP populations in the Galactic bulge and the globular cluster M15, which is used to derive constraints on $\eta_e/\eta_\gamma$ from \textsc{Fermi}-LAT observations and projected \textsc{Ctao} sensitivities toward the Galactic bulge, as well as from \textsc{Magic} observations of M15.

\subsection{Galactic bulge observations}
\label{subsec:GCflux}

The injection rate of $e^\pm$ or $\gamma$-rays from MSPs at an energy $E$ and position $\vec{x}$ is written as follows:

\begin{equation}
    \label{eq:source}
    q_i(\vec{x},E_i) = N_{\rm{MSP}}
    \left(\frac{\rho_{\rm{bulge}}(\vec{x})}{M_{\rm{bulge}}}\right) \int dL_i\,\frac{dN_i}{dL_i}\,\frac{d^2N_i}{dE_idt}(E_i;L_i)
\end{equation}
where $i=\{\gamma,e\}$, $N_{\rm{MSP}}$ is the number of MSPs in the bulge, $\rho_{\rm{bulge}}$ is bulge profile. $M_{\rm{bulge}} = \iiint d^3\vec{x}\,\rho_{\rm{bulge}}(\vec{x})$ is the total mass of the bulge, $dN/dL$ is the MSP luminosity distribution and $d^2N_i/(dE_idt)$ is the injection spectrum of $e^\pm$ or $\gamma$-rays of an individual MSP given its $e^\pm$ or $\gamma$-ray luminosity $L_i$, which is normalized so that
\begin{equation}
    \label{eq:norm}
    \int dE_i\,E_i\,\frac{d^2N_i}{dE_idt}(E_i;L_i) = L_i\,.
\end{equation}
The injection spectrum  $d^2N_i/(dE_idt)$ is usually parametrized as power-law of index $\Gamma$ with a cutoff $E_{\rm{cut}}$, with normalization $A$
\begin{equation}
    \label{eq:injecspec}
    \frac{d^2N_i}{dE_idt}(E_i;L_i) = A(L_i)\,E_i^{-\Gamma}e^{-E_i/E_{\rm{cut}}},
\end{equation}
where for $e^\pm$, $\Gamma$ lies between 1.5 and 2.5 and $E_{\rm{cut}} \sim$ tens of TeV~\cite{Song:2019nrx}, while for prompt $\gamma$-rays, $\Gamma \approx 1.5$ and $E_{\rm{cut}} \approx 3.3$ GeV~\cite{Petrovic:2014uda}. 

The total luminosity of photons emitted by the MSPs in the bulge $L_{\gamma,{\rm{tot}}}$ can be related to their injection rate:
\begin{equation}
    \label{eq:Lgamma}
    L_{\gamma,{\rm{tot}}} = \iiint d^3\vec{x} \int_0^\infty dE_\gamma\,E_\gamma\,q_\gamma(\vec{x},E_\gamma) = N_{\rm{MSP}}\langle L_\gamma\rangle
\end{equation}
where $\langle L_\gamma\rangle \equiv \int dL_\gamma\,L_\gamma\,dN_\gamma/dL_\gamma$ is the average MSP $\gamma$-ray luminosity. If $\dot{E}_{\rm{tot}}$ is the sum of the spin-down power over all MSPs in the bulge, and each MSP has a fraction $\eta_\gamma$ of their spin-down power converted into photons, then 
\begin{equation}
    L_{\gamma,{\rm{tot}}} \equiv \eta_\gamma \dot{E}_{\rm{tot}}\;.
\end{equation}
The same reasoning applies to the total $e^\pm$ luminosity $L_e$:
\begin{equation}
    \begin{split}
        L_{e,\rm{tot}} &\equiv \eta_e \dot{E}_{\rm{tot}} = \frac{\eta_e}{\eta_\gamma} L_{\gamma,\rm{tot}} \\
        &= N_{\rm{MSP}}\langle L_e\rangle \quad \textrm{(similarly to Eq.~\ref{eq:Lgamma})}
    \end{split}
    \label{eq:Le}
\end{equation}
where $\langle L_e\rangle \equiv \int dL_e\,L_e\,dN_e/dL_e$ is the average MSP luminosity in $e^\pm$.

As for the prediction of the differential photon flux measured at Earth over a region of interest $\Delta\Omega$, we have to take into account two components. The first is the prompt emission of $\gamma$-rays directly emitted by the MSPs in the bulge:
\begin{equation}
    \label{eq:prompt}
    \frac{d\Phi_\gamma^{\rm{prompt}}}{dE_\gamma} = \frac{1}{4\pi}\int_{\Delta\Omega}d\Omega\int_{\rm{l.o.s.}} ds\,q_\gamma(\vec{x}(s,\Omega),E_\gamma)\;,
\end{equation}
and using Eqs.~\ref{eq:source}, \ref{eq:norm}, and \ref{eq:injecspec} we obtain
\begin{equation}
    \label{eq:promptsimpAppx}
    \frac{d\Phi_\gamma^{\rm{prompt}}}{dE_\gamma} =\frac{\mathcal{D}}{4\pi}\frac{L_{\gamma,\rm{tot}}}{L_{\gamma}}\,\frac{d^2N_\gamma}{dE_\gamma dt}(E_\gamma;L_\gamma)\;,
\end{equation}
where
\begin{equation}
    \mathcal{D} = \int_{\Delta\Omega}d\Omega\int_{\rm{l.o.s.}} ds\,\left(\frac{\rho_{\rm{bulge}}(\vec{x}(s,\Omega))}{M_{\rm{bulge}}}\right)
\end{equation}
is analogous to the $D$-factor in DM-related studies. Eq.~\ref{eq:promptsimpAppx} is actually independent on $L_\gamma$ since $d^2N_\gamma/(dE_\gamma dt)\propto L_\gamma$.

The second component is related to secondary emissions of photons, notably from the inverse-Compton scattering (ICS) of interstellar radiation field (ISRF) photons in the bulge by highly energetic MSP-produced $e^\pm$. The differential photon flux from ICS reads:
\begin{equation}
    \label{eq:ICS}
    \frac{d\Phi_\gamma^{\rm{ICS}}}{dE_\gamma} = \frac{1}{4\pi E_\gamma}\int_{\Delta\Omega}d\Omega\int_{\rm{l.o.s.}} ds
    \int_{m_e}^\infty dE_e\,\mathcal{N}_e(\vec{x}(s,\Omega),E_e)\mathcal{P}_{\rm{ICS}}(\vec{x}(s,\Omega),E_\gamma,E_e)
\end{equation}
where $\mathcal{P}_{\rm{ICS}}$ is the ICS power radiated by a single electron~\cite{Cirelli:2009vg}, $\mathcal{N}_e$ is the propagated $e^\pm$ density, which is the solution of the diffusion-loss equation
\begin{equation}
    -\vec\nabla \cdot(D(\vec x, E_e)\vec\nabla \mathcal{N}_e)-\frac{\partial}{\partial E_e}(b(\vec{x}, E_e)\mathcal{N}_e)=q_e(\vec{x}, E_e)\;.
\end{equation}
Near the GC the inner Galaxy energy losses dominates over spatial diffusion~\cite{DiMauro:2021qcf}, therefore $\mathcal{N}_e$ simplifies drastically:
\begin{equation}
    \mathcal{N}_e(\vec{x},E_e) = \frac{1}{b(\vec{x},E_e)}\int_{E_e}^\infty dE'_e\,q_e(\vec{x},E'_e)\;,
\end{equation}
where $b$ encodes the energy losses. In this manner, and using Eqs.~\ref{eq:source}, \ref{eq:norm}, \ref{eq:injecspec}, and \ref{eq:Le}, Eq.~\ref{eq:ICS} can be rewritten:
\begin{equation}
    \label{eq:ICSsimpAppx}
    \frac{d\Phi_\gamma^{\rm{ICS}}}{dE_\gamma} = \frac{\mathcal{D}}{4\pi E_\gamma}\frac{\eta_e}{\eta_\gamma}\frac{L_{\gamma,\rm{tot}}}{L_e}\,\int_{E_e^{\rm{min}}}^\infty dE_e\frac{\mathcal{P}_{\rm{ICS}}(E_\gamma,E_e)}{b(E_e)}\int_{E_e}^\infty dE_e'\,\frac{d^2N_e}{dE_e'dt}(E'_e;L_e)\,
\end{equation}
where we assume the homogeneity of energy losses and radiating power. Like Eq.~\ref{eq:promptsimpAppx}, Eq.~\ref{eq:ICSsimpAppx} is independent on $L_e$ since $d^2N_e/(dE'dt)\propto L_e$.

In the end, we find that $L_{\gamma,\rm{tot}}$ controls the normalization of the total photon flux while $\eta_e/\eta_\gamma$ the normalization of the ICS component. We can further constrain these two parameters using the measured spectrum of the GCE. We can relate the measured luminosity of the GCE $L_{\rm{GCE}}^{\rm meas}$ with the predicted spectrum
\begin{equation}
    \label{eq:LGCEAppx}
    L_{\rm{GCE}}^{\rm meas} = 4\pi r_\odot^2 \int_{E_{\gamma,\rm{GCE}}^{\rm{min}}}^{E_{\gamma,\rm{GCE}}^{\rm{max}}} dE_\gamma\;E_\gamma\left(\frac{d\Phi_\gamma^{\rm{prompt}}}{dE_\gamma}+\frac{d\Phi_\gamma^{\rm{ICS}}}{dE_\gamma}\right)\;,
\end{equation}
which directly depend on $L_{\gamma,\rm{tot}}$ and $\eta_e/\eta_\gamma$ as shown in Eqs.~\ref{eq:promptsimpAppx} and~\ref{eq:ICSsimpAppx}. Combining Eqs.~\ref{eq:promptsimpAppx}, \ref{eq:ICSsimpAppx} and \ref{eq:LGCEAppx}, the total photon flux is written
\begin{equation}
    \label{eq:fluxtotsimp1}
    \frac{d\Phi_\gamma^{\rm tot}}{dE_\gamma} = \frac{L_{\rm GCE}^{\rm meas}}{4\pi r_{\odot}^2} \frac{F(E_\gamma)}{\int_{\Delta E_{\rm GCE}}dE_\gamma\,E_\gamma F(E_\gamma)}\;,
\end{equation}
where 
\begin{equation}
    \label{eq:fluxtotsimp2}
    F(E_\gamma)=\frac{1}{L_\gamma}\frac{d^2N_\gamma}{dE_\gamma dt}(E_\gamma;L_\gamma)+\frac{\eta_e}{\eta_\gamma}\frac{1}{E_\gamma}\int_{E_e^{\rm min}}^\infty dE_e \frac{\mathcal{P}_{\rm ICS}(E_\gamma,E_e)}{b(E_e)}\int_{E_e}^\infty dE_e'\,\frac{1}{L_e}\frac{d^2N_e}{dE_e'dt}(E'_e;L_e)\;.
\end{equation}
Given that $d^2N_i/(dE_idt)\propto L_i$, these two equations show that the free parameters are the ones setting the photon and electron injection spectra ($\Gamma_i$ and $E_{c,i}$, both encoded in $d^2N_i/(dE_i dt)$, for $i\in\{\gamma,e\}$), and the ratio of efficiencies $\eta_e/\eta_\gamma$.

\subsection{Globular cluster M15 observations}
\label{subsec:M15flux}

We extend the previous subsection to the study of the globular cluster M15, for which we will consider as a point source. In this case, the electron source term is written
\begin{equation}
    \begin{split}
        q_e(\vec{x},E_e) &=N_{\rm MSP}\int dL_e\,\frac{dN_e}{dL_e}\,\frac{d^2N_e}{dE_edt}(E_e;L_e)\,\delta^{(3)}(\vec{x}-\vec{x}_{\rm M15})  \\
        &= \frac{\eta_e}{\eta_\gamma}\frac{L_{\gamma,\rm tot}}{L_e}\frac{d^2N_e}{dE_edt}(E_e;L_e)\,\delta^{(3)}(\vec{x}-\vec{x}_{\rm M15}) \;.
    \end{split}
\end{equation}
Given that the \textsc{Magic} energy range extends from $50$ to $2\times10^5$~GeV, the predicted photon flux in this regime is dominated by ICS. In addition, the \textsc{Magic} observations cover a region more than an order of magnitude larger than the tidal radius of M15 ($\sim30$~pc). As a result, for any realistic diffusion scenario of MSP-injected $e^\pm$, the predicted spectrum of M15 remains effectively unchanged. We therefore adopt the no-diffusion approximation in our calculation. Therefore, we can write the ICS flux from M15, using the fact that $\delta^{(3)}(\vec{x}-\vec{x}_{\rm M15})=\delta(s-d)\delta(\theta-\theta_{\rm M15})\delta(\varphi-\varphi_{\rm M15})/(s^2\sin\theta)$, and we obtain
\begin{equation}
    \frac{d\Phi_\gamma^{\rm{ICS}}}{dE_\gamma} = \frac{1}{4\pi d^2E_\gamma}\frac{\eta_e}{\eta_\gamma}\frac{L_{\gamma,\rm{tot}}}{L_e}\,\int_{E_e^{\rm{min}}}^\infty dE_e\frac{\mathcal{P}_{\rm{ICS}}(E_\gamma,E_e)}{b(E_e)}\int_{E_e}^\infty dE_e'\,\frac{d^2N_e}{dE_e'dt}(E'_e;L_e)\;,
\end{equation}
where $d\simeq10.4$ kpc~\cite{1996AJ....112.1487H} is the distance between Earth and M15. Assuming that the photon flux from ICS is negligible to prompt photons (it constitutes around 3\% of the total flux in this range in the case of the GB), the total prompt photon luminosity of M15 is approximately related to the flux above 100 MeV measured by \textsc{Fermi}-LAT $L_{\gamma,\rm{tot}}\simeq4\pi d^2G_{\gamma}(> 100\,\rm{MeV})$ and therefore simplifies even more the previous equation
\begin{equation}
    \frac{d\Phi_\gamma^{\rm{ICS}}}{dE_\gamma} = \frac{G_{\gamma}(> 100\,\rm{MeV})}{E_\gamma}\frac{\eta_e}{\eta_\gamma}\,\int_{E_e^{\rm{min}}}^\infty dE_e\frac{\mathcal{P}_{\rm{ICS}}(E_\gamma,E_e)}{b(E_e)}\int_{E_e}^\infty dE_e'\,\frac{1}{L_e}\frac{d^2N_e}{dE_e'dt}(E'_e;L_e)\;,
\end{equation}
with $G_{\gamma}(> 100\,\rm{MeV}) = (3.57 \pm 0.58) \times 10^{-12}$ erg cm$^{-2}$ s$^{-1}$.

The astrophysical environment of M15 remains rather uncertain. Pulsars are known to generate strongly magnetized winds, and interactions among winds from a population of pulsars in a globular cluster can produce magnetic fields ranging from a few to several tens of $\mu$G~\cite{Bednarek:2007nn}. Following Ref.~\cite{MAGIC:2019aof}, we therefore consider magnetic-field strengths in the range $1$--$30~\mu$G. These magnetic fields contribute to the energy losses of MSP-injected $e^\pm$ through synchrotron radiation. In addition, the optical radiation field inside M15 is not precisely known, owing to uncertainties in the stellar population of the cluster. For the spatial distribution of stars, we adopt a Michie–King profile~\cite{1963MNRAS.125..127M,1963MNRAS.126..269M,1963MNRAS.126..331M,1963MNRAS.126..499M,Kuranov:2006kw}, while the photon spectrum is modeled as a blackbody with temperature $5000$~K. Using the measured stellar luminosity of M15, we estimate the average energy density of stellar photons within the tidal radius to be $\langle u_{\rm opt}\rangle \simeq 0.43~\mathrm{eV\,cm^{-3}}$. We emphasize, however, that including this optical component significantly affects the predicted ICS flux only for electron injection scenarios dominated by relatively low-energy $e^\pm$ (with cutoff energies below $\sim0.5$~TeV). At higher energies, the ICS emission is instead dominated by scattering on CMB photons.

\section{Previous constraints}
\label{app:Previous_constraints}
Globular clusters have long been considered promising targets for searches of high-energy emission associated with populations of millisecond pulsars (MSPs). These systems host large numbers of old neutron stars formed through stellar interactions in their dense cores, leading to the formation of MSPs that can efficiently accelerate relativistic particles. Similarly to what happens in the GC, the electrons and positrons injected by these objects may interact with the intense stellar radiation fields present in globular clusters and produce gamma rays through IC scattering. As a result, globular clusters have been proposed as potential sources of high-energy gamma-ray emission.

The interest in these objects increased significantly after the detection of several globular clusters in the GeV range by the \textsc{Fermi}-LAT, including well-known systems such as 47~Tucanae and Terzan~5~\cite{Abdo:2009rjc, Kong2010_Terzan5, Fermi-LAT:2010gpc}. These detections are generally interpreted as the cumulative emission from a population of MSPs within the clusters. Motivated by these observations, several theoretical works have investigated the possibility that relativistic leptons injected by MSPs could also produce detectable emission at higher energies, in particular in the TeV range, making such objects interesting targets to study MSPs particle injections. 
A number of observational campaigns have therefore targeted globular clusters with ground-based Cherenkov telescopes. Experiments such as \textsc{Hess}, \textsc{Magic}, and \textsc{Veritas} have performed dedicated searches for TeV gamma-ray emission from several clusters, including Terzan~5~\cite{Abramowski2011Terzan5}, 47~Tucanae~\cite{Abramowski2013GCs}, M13~\cite{Anderhub2009M13}, and M15~\cite{McCutcheon2009VERITAS, MAGIC:2019aof}. While most observations have resulted in non-detections, they have provided valuable upper limits on the gamma-ray flux and have been used to constrain the efficiency with which MSPs inject relativistic leptons into the cluster environment. 


Using about 165 hours of observations of the globular cluster M15, \textsc{Magic} used the non-detection of M15 to set the strongest constraints on the efficiency with which MSPs convert their rotational energy loss into relativistic leptons. 
Depending on the assumed injection scenario, the \textsc{Magic} limits imply $\eta_e/\eta_{\gamma} \gtrsim 0.1$, which is somewhat lower than the efficiencies typically expected in pulsar environments. 
However, these limits are subject to substantial astrophysical uncertainties. A major source of uncertainty is the magnetic field strength within the cluster, typically assumed to lie in the range $B \sim 1$--$10\,\mu\mathrm{G}$. 
Another key uncertainty concerns particle transport. If leptons remain confined and diffuse slowly within the cluster, efficient IC emission is expected and the \textsc{Magic} limits become restrictive. Conversely, if advection by cluster winds allows particles to escape rapidly, the gamma-ray signal can be strongly suppressed, in which case the non-detection does not necessarily imply a low injection efficiency.
Additional uncertainties arise from the poorly known properties of the MSP population and the cluster radiation environment. Only a handful of MSPs are currently detected in M15, but the true population could be considerably larger, affecting the total injected power. 

\section{Prospects with CTAO}
\label{app:CTA}

The upcoming \textsc{Ctao}, with its unprecedented sensitivity, angular resolution, and energy coverage in the TeV range, offers significant opportunities to detect or constrain high‑energy emission associated with MSP populations in the GC region. While the \textsc{Ctao} consortium’s sensitivity studies to DM signals focus primarily on annihilation signatures in the inner Galaxy~\cite{CTA:2020qlo}, it is precisely this deep and high‑resolution survey of the inner few degrees that also enhances prospects for studying other diffuse and unresolved components such as MSP‑induced inverse Compton emission. Independent forecasts for \textsc{Ctao}’s sensitivity to MSP populations indicate that, under favorable assumptions about the injected $e^\pm$ spectrum (e.g.\ harder than \(E^{-2}\)), \textsc{Ctao} could robustly detect the TeV‑scale IC signal from a bulge MSP population sufficient to explain the GeV excess; even in more conservative scenarios, \textsc{Ctao} can place strong limits on MSP injection efficiencies and emission models~\cite{Macias:2021boz}.

Furthermore, \textsc{Ctao}’s ability to resolve spectral and spatial features in the high‑energy gamma‑ray sky provides a powerful discriminator between MSP‑driven models and alternative interpretations, including DM annihilation. By comparing the expected TeV emission morphology and spectrum from unresolved MSP populations with \textsc{Ctao} observations, it will be possible to either confirm the presence of a high‑energy component associated with MSPs or constrain their contribution to the GCE, thereby improving our understanding of the underlying source population and the particle acceleration processes at play in the inner Galaxy.

\section{Background Modeling of the Galactic Center and the Galactic Center Excess}
\label{appx:IEMs}

The $\gamma$-ray emission from the inner Galaxy is overwhelmingly dominated by Galactic interstellar emission, which accounts for approximately 90\% of the observed flux, while the Galactic Center Excess (GCE) contributes only 5--10\%. As a consequence, any attempt to characterize the spectrum or morphology of the GCE is inherently limited by the accuracy of the interstellar emission model (IEM).

This dominance introduces several critical challenges. First, systematic uncertainties in the IEM directly propagate into the inferred properties of the GCE. Second, degeneracies between diffuse components---such as inverse-Compton (IC), bremsstrahlung, and $\pi^0$-decay emission---complicate the separation of distinct physical contributions. Third, mismodeling of the Galactic plane, where emission is both bright and spatially structured, can produce artificial residuals and inflate statistical significances, particularly at low energies.

To mitigate these effects, analyses typically explore a range of physically motivated IEMs and allow for flexibility in their normalization and spectral shapes. This approach is essential to assess the robustness of any detected excess against diffuse-modeling assumptions and to avoid misinterpreting background mismodeling as a genuine signal.

All IEMs considered in this work include the following components: (i) Inverse Compton (IC) emission from cosmic-ray electrons scattering on the interstellar radiation field (CMB, infrared, and starlight), (ii) bremsstrahlung from cosmic-ray leptons interacting with interstellar gas, (iii) $\pi^0$-decay from hadronic cosmic-ray interactions, (iv) large-scale structures such as the \textsc{Fermi} bubbles and Loop I, and (v) isotropic, solar, and lunar contributions. The IEMs we adopted in this work have different characteristics, which are described below.



\begin{itemize}
    \item \textbf{Galp21 (SA0, SA50, SA100)}~\cite{Porter:2017vaa}: Public \texttt{GALPROP} models with different CR source distributions (axisymmetric, mixed disk + spiral arms, and pure spiral arms). Each of these models includes templates for the \textsc{Fermi} bubbles, bremsstrahlung, $\pi^0$ decay, and three IC components (CMB, infrared, and starlight). In addition, they are ring-decomposed for the ICS emission, and gas-correlated components yield approximately 20 templates, providing strong spatial flexibility.

    \item \textbf{Macias18}~\cite{Macias:2016nev}: Includes improved dust-based gas corrections and explicit dust residual templates. It is also ring-decomposed into 6 IC and 4 gas rings (19 templates total) and is often combined with stellar-bulge templates to trace a putative MSP population.

    \item \textbf{Pohl22}~\cite{Pohl:2022nnd}: An update of Macias18, implementing a revised reconstruction of the inner-Galaxy H\textsc{i} distribution. The overall ring/template structure remains the same, providing a targeted probe of gas-modeling systematics.
\end{itemize}

The values of the measured GCE lumonisity $L_{\rm GCE}^{\rm meas}$ for each of the considered IEMs are reported in Tab.~\ref{tab:LGCE}. The flux extracted in each dataset was integrated over the photon energy range where \textsc{Fermi}-LAT reported a measurement. These energy ranges are also reported in Tab.~\ref{tab:LGCE}. The value of the Galactocentric distance $r_\odot$ is also crucial to compute $L_{\rm GCE}^{\rm meas}$, and depends on the spatial template used to extract the datasets. The Galp21 and Pohl22 datasets were extracted using the non-parametric template of Ref.~\cite{Coleman:2019kax}, while Macias18 was extracted using the S-model~\cite{Freudenreich:1997bx}. The values of $r_\odot$ used in these templates are reported in Tab.~\ref{tab:LGCE} as well.

\begin{table}[t]
    \centering
    \begin{tabular}{|c|c|c|c|}
        \cline{2-4}
        \multicolumn{1}{c|}{} & $L_{\rm GCE}^{\rm meas}$ [$\times10^{36}$ erg/s] & $\Delta E_{\rm GCE}$ [GeV] & $r_\odot$ [kpc] \\
        \hline
        Macias18 & $6.93\pm0.31$ & $0.5-10.0$ & $8.5$ \\
        \hline
        GalpSA0 & $5.41\pm0.29$ & $0.5-17.8$ & $7.9$ \\
        \hline
        GalpS50 & $5.34\pm0.25$ & $0.5-17.8$ & $7.9$ \\
        \hline
        GalpSA100 & $7.40\pm0.33$ & $0.5-13.3$ & $7.9$ \\
        \hline
        Pohl22 & $5.27\pm0.24$ & $0.5-10.0$ & $7.9$ \\
        \hline
    \end{tabular}
    \caption{Measured values of the luminosity of the GCE $L_{\rm GCE}^{\rm meas}$, the associated detection energy range $\Delta E_{\rm GCE}$, and the Galactocentric distance $r_\odot$ used for the extraction of the five considered \textsc{Fermi}-LAT datasets.}
    \label{tab:LGCE}
\end{table}

\section{Additional results}
\label{appx:add_results}

This section showcases additional results of our study, including the impact of the systematic uncertainties in the upper limits on $\eta_e/\eta_\gamma$ from \textsc{Fermi}-LAT observations of the GC, comparisons between the flux prediction and the \textsc{Ctao} sensitivity to the GC, as well as M15 observations. Finally, we show additional upper limits on $\eta_e/\eta_\gamma$ for different $e^\pm$ injection cut-off energies $E_{c,e}$.

\subsection{Systematic uncertainties}

\begin{figure}[t]
    \centering
    \includegraphics[width=0.49\linewidth]{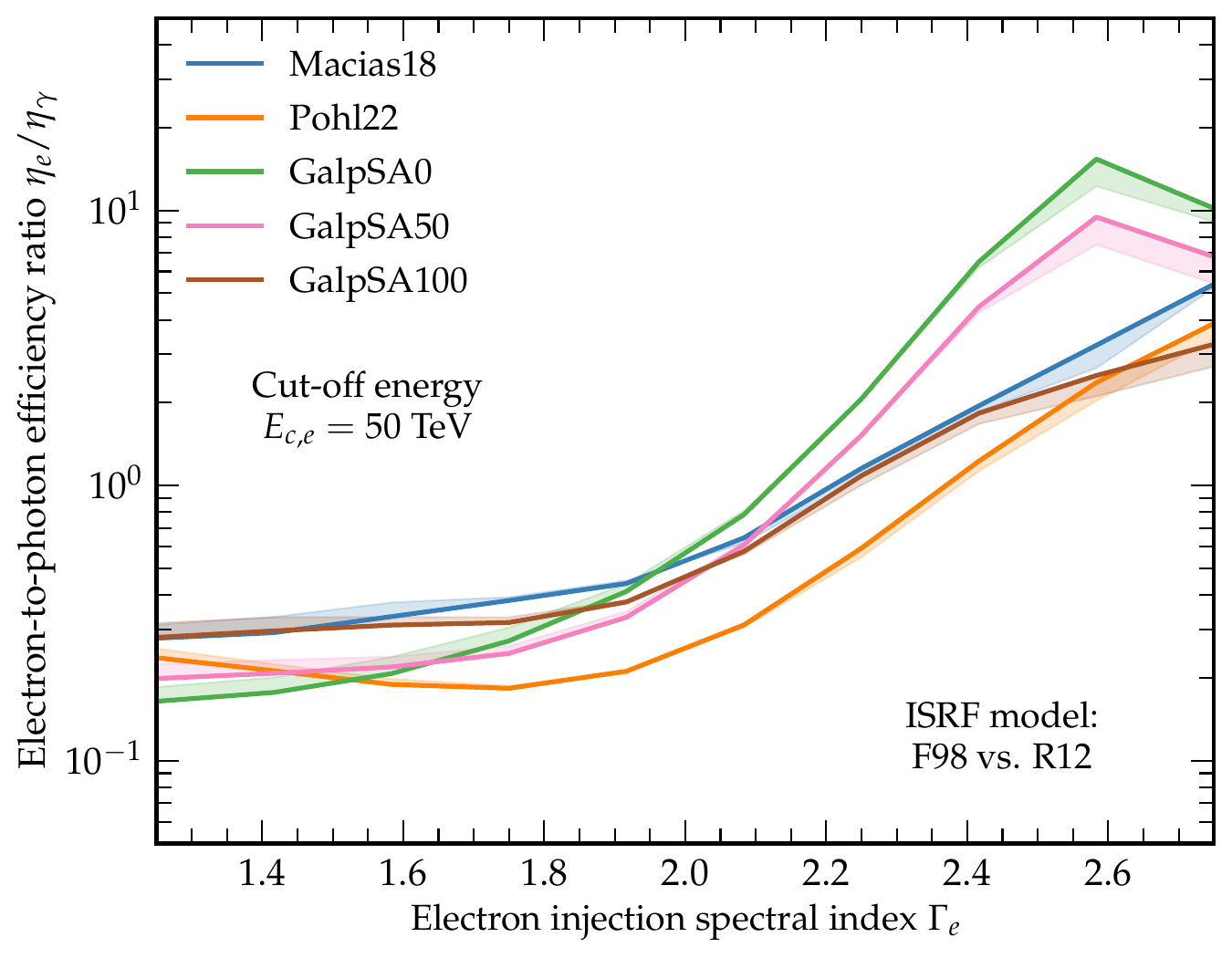}
    \includegraphics[width=0.49\linewidth]{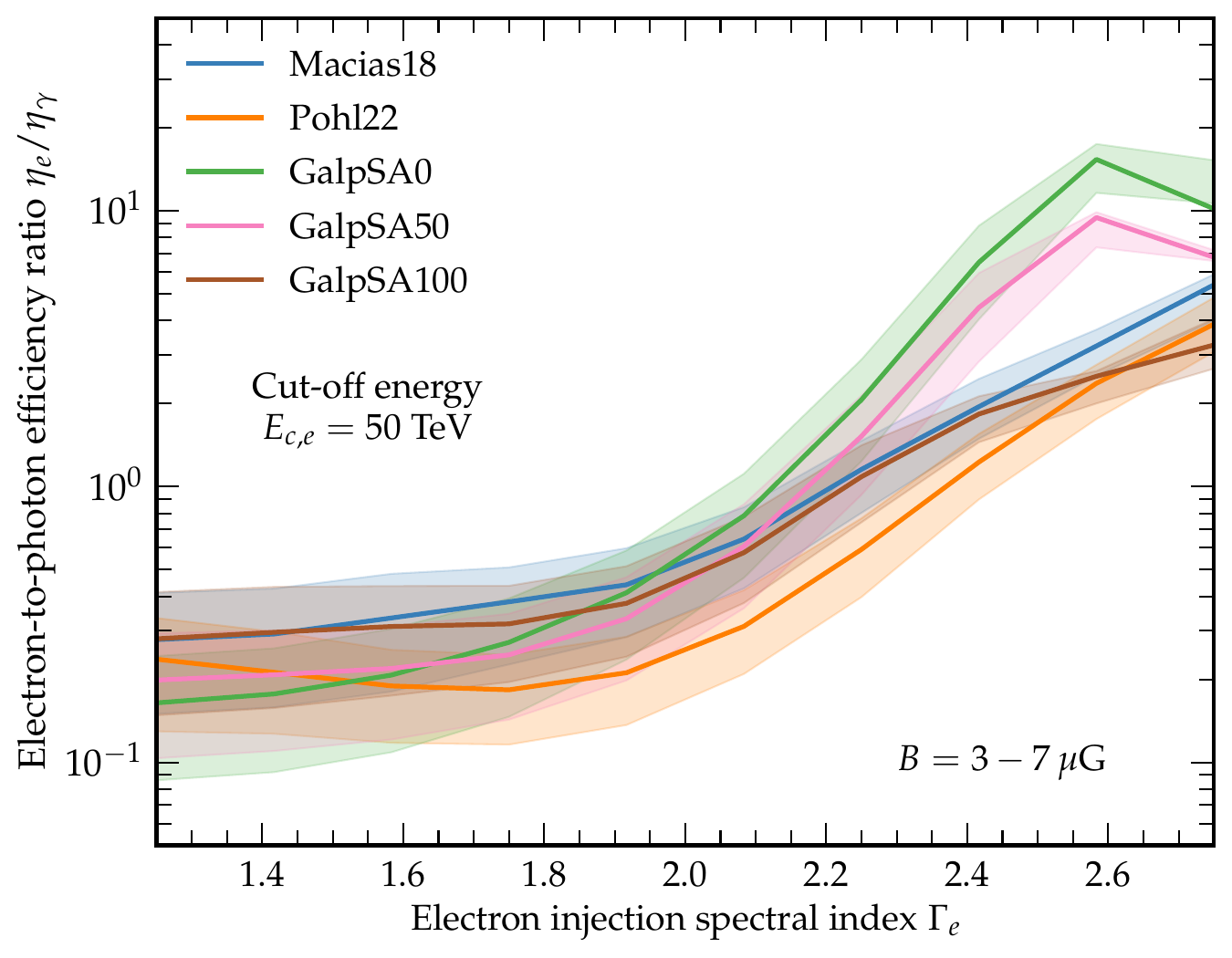}
    \includegraphics[width=0.49\linewidth]{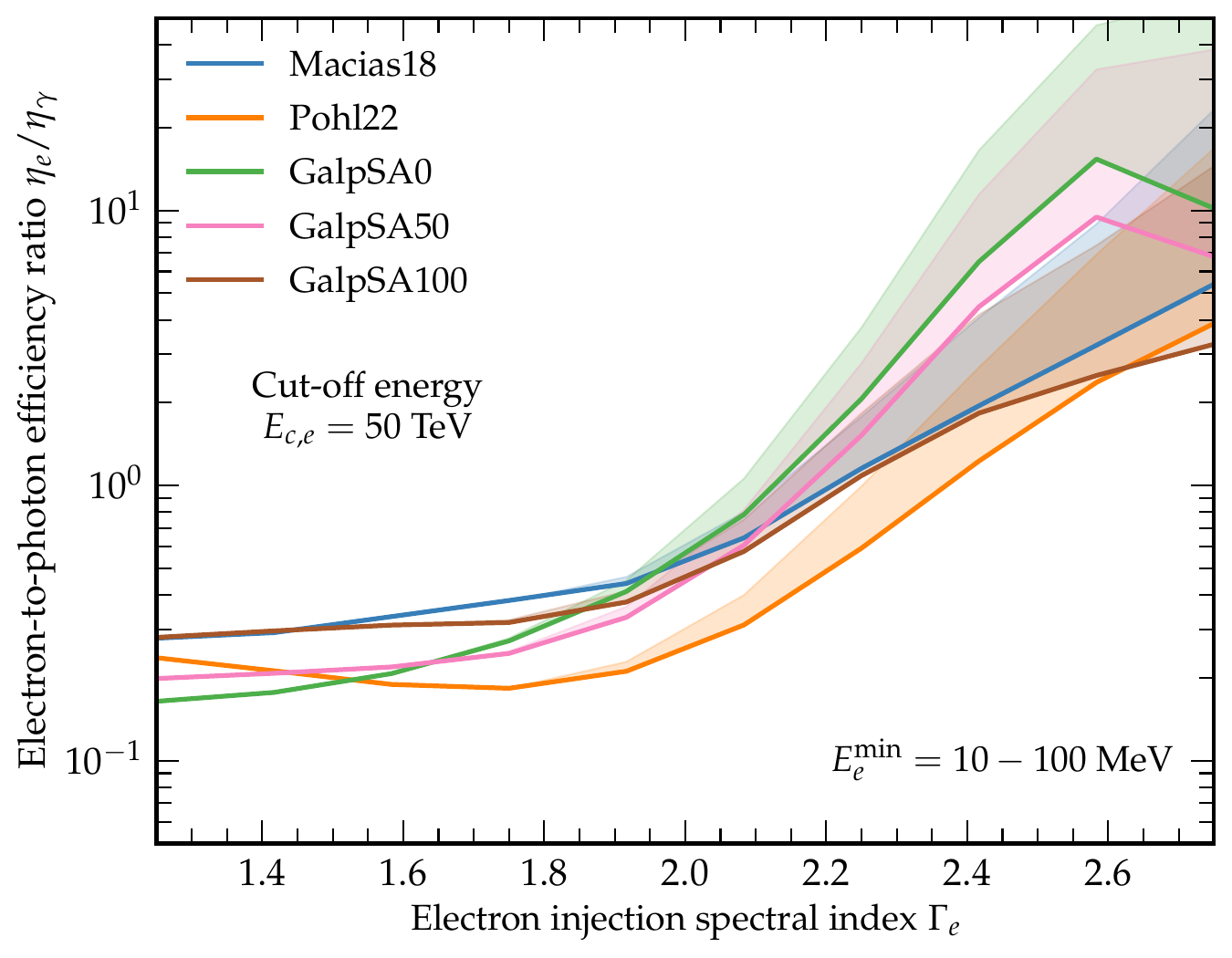}
    \includegraphics[width=0.49\linewidth]{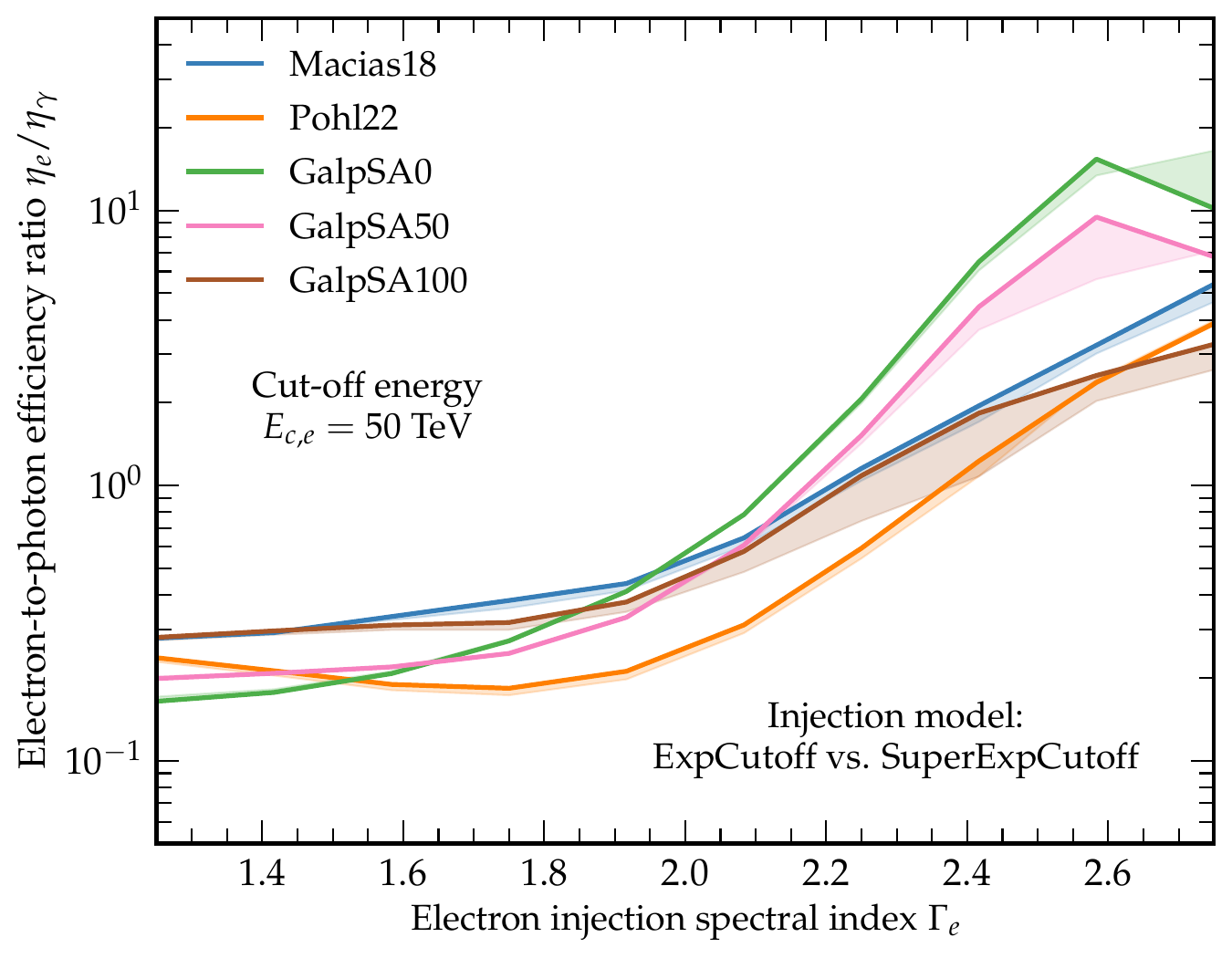}
    \caption{\textit{Top Left:} Upper limits on $\eta_e/\eta_\gamma$ from the different IEMs (\textsc{Fermi}-LAT datasets) used in our analysis. The colored bands represent a variation in the starlight and infrared photon field model (F98 and R12) adopted in our prediction. \textit{Top Right:} Same as the previous panel, but the bands showcase the impact of adopting different average magnetic field values (from 3 to 7 $\mu$G). \textit{Bottom Left:} Same as the other panels, but the bands showcase the impact of adopting different minimal MSP-injected $e^\pm$ energies $E_e^{\rm min}$ (from 10 to 100 MeV). \textit{Bottom Right:} Same as the other panels, but the bands showcase the impact of adopting a power-law with a super-exponential cutoff for the prompt $\gamma$-ray injection spectrum. In all of the panels, the solid line represents the results when using F98 and $5.4\mu$G, $E_e^{\rm{min}}=100$ MeV and a regular power-law with exponential cutoff for the prompt MSP-injected $\gamma$-rays.}
    \label{fig:unc_UL}
\end{figure}

In our analysis with \textsc{Fermi}-LAT observations of the GB, we consider three sources of systematic uncertainties: the choice of the IEM template used to extract the \textsc{Fermi}-LAT dataset (Macias18, GalpSA0, GalpSA50, GalpSA100 and Pohl22), the choice of the starlight and infrared photon field model (F98 and R12), and the average value of the Galactic magnetic field which impacts the $e^\pm$ energy losses from synchrotron radiation (3 to 7 $\mu$G). The top panels of Fig.~\ref{fig:unc_UL} shows the impact of these systematic uncertainties in the upper limits on $\eta_e/\eta_\gamma$ we compute from \textsc{Fermi}-LAT observations. Moreover, adopting different values of the minimal energy of MSP-injected $e^\pm$ can drastically impact these bounds for harder $e^\pm$ injection spectra ($\Gamma_e\gtrsim2$). In our analysis, we adopted $E_e^{\rm min}=100$ MeV as a benchmark. We show the impact on the derived upper limits on $\eta_e/\eta_\gamma$ when choosing $E_e^{\rm min}=10$ MeV on the bottom left panel of Fig.~\ref{fig:unc_UL}. Finally, one could in principle adopt a more general injection spectrum of prompt $\gamma$-rays to fit better the spectrum of the GCE measured by \textsc{Fermi}-LAT, e.g. a power law with a super-exponential cutoff $d^2N_\gamma/(dE_\gamma dt) \propto E_\gamma^{-\Gamma}\exp(-(E_\gamma/E_c)^\beta)$, where $\beta$ caracterizes the strength of the cutoff. We however find that our fitting procedure prefers a slightly sub-exponential cutoff ($\beta\sim0.85$), therefore increasing the prompt $\gamma$-ray flux at higher energies, hence strengthening (by a few percents) the upper limits on $\eta_e/\eta_\gamma$. We show this impact on the bottom right panel of Fig.~\ref{fig:unc_UL}.

The two aforementioned uncertainty sources impact the upper limits derived from the \textsc{Ctao} sensitivity, as they directly impact the prediction of the photon flux from ICS. Additionally, the choice of the IEM also impacts these upper limits, as the value of $L_{\gamma,{\rm tot}}$ (although corrected for the different ROI), is derived with \textsc{Fermi}-LAT observations.

For the study of the M15 globular cluster, the dominant source of systematic uncertainty arises from the assumed value of the internal magnetic field. In our analysis, we adopt the range $1$--$30~\mu$G, following Ref.~\cite{MAGIC:2019aof}. A secondary source of systematic uncertainty originates from the determination of $L_{\gamma,\rm tot}$, which is inferred from the \textsc{Fermi}-LAT measurement of the photon flux above 100~MeV: $G_{\gamma}(> 100\,\rm{MeV}) = (3.57 \pm 0.58) \times 10^{-12}$ erg cm$^{-2}$ s$^{-1}$.

\subsection{Flux predictions versus CTAO sensitivity and MAGIC observations}

\begin{figure}[t]
    \centering
    \includegraphics[width=0.49\linewidth]{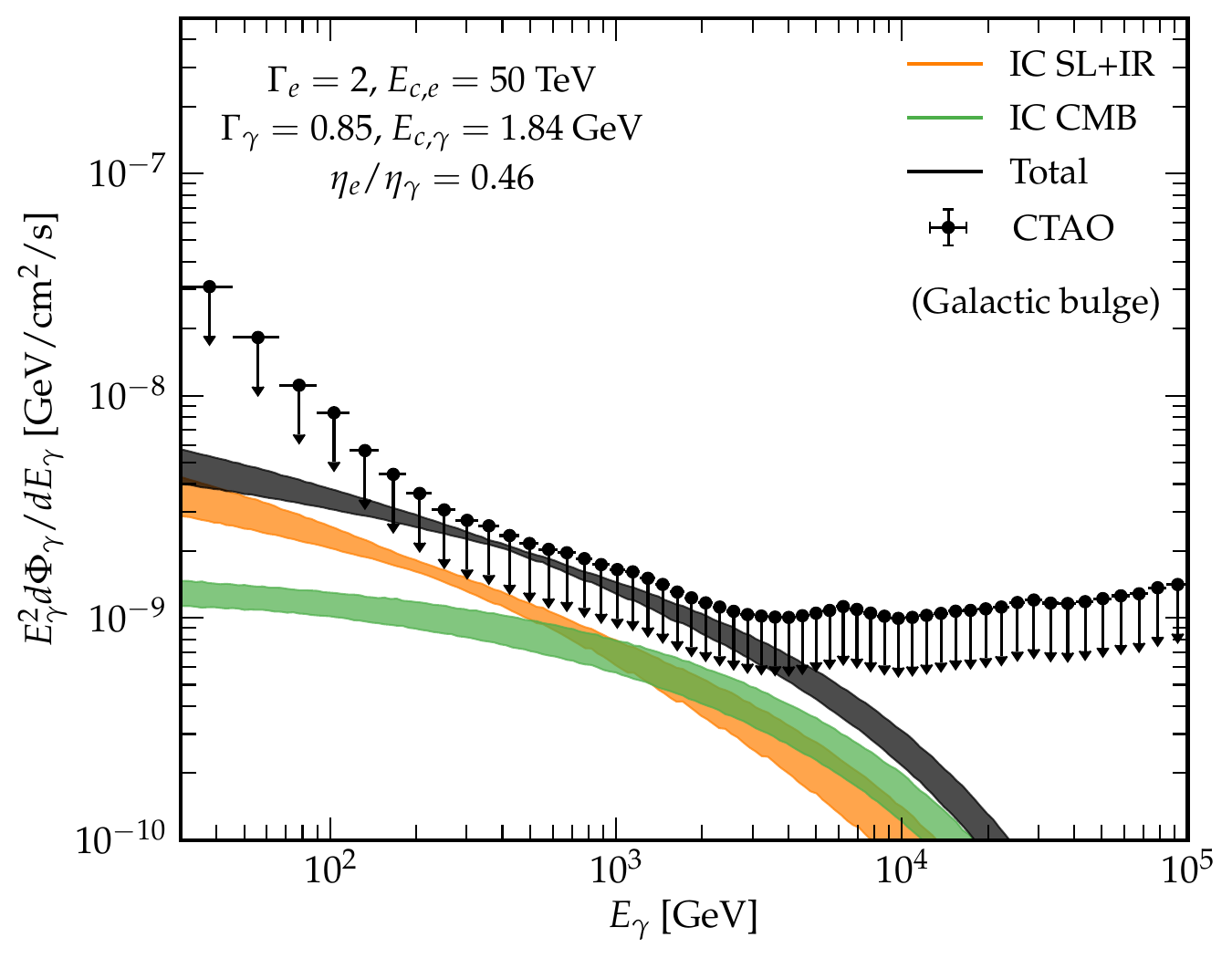}
    \includegraphics[width=0.49\linewidth]{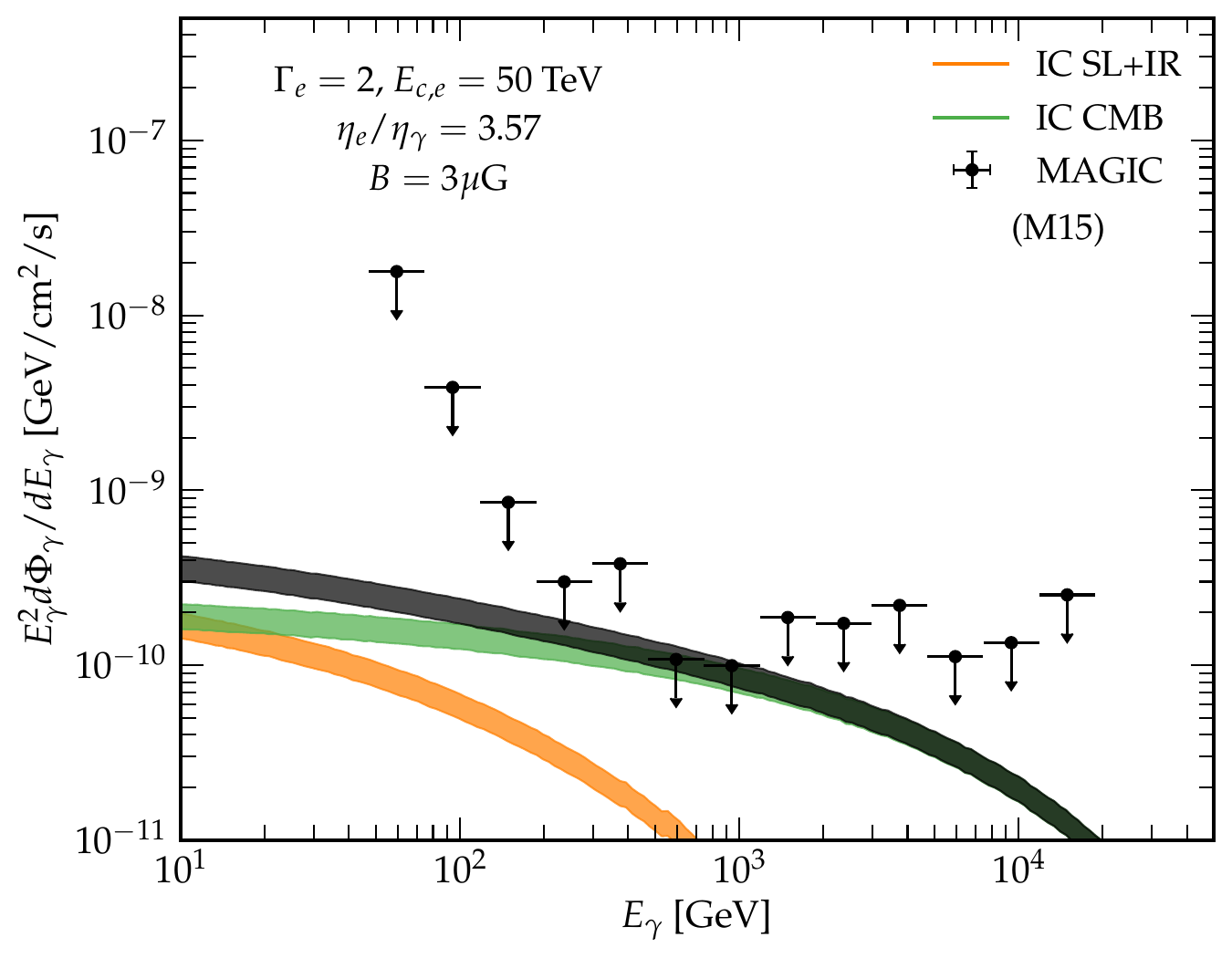}
    \caption{\textit{Left:} Predicted total photon flux (black band) from MSPs in the GB, fitted to the \textsc{Ctao} sensitivity (black points). The fitted parameters are the same as in Fig.~\ref{fig:FitExample}, except for the efficiency ratio $\eta_e/\eta_\gamma$. The orange band shows the systematic uncertainties related to the choice of the starlight and infrared photon field model (F98 and R12) and the magnetic field (3 to 7 $\mu$G), while the green band represents the ICS contribution on the CMB. \textit{Right:} Predicted total photon flux (green) from MSPs in the M15 globular cluster, fitted to the observations from \textsc{Magic} (black points). The prediction has been made for $B=1\,\mu$G, and the bands represents the uncertainties coming from the \textsc{Fermi}-LAT measurement of $G_{\gamma}(> 100\,\rm{MeV})$.}
    \label{fig:FitExampleCTAO}
\end{figure}

Fig.~\ref{fig:FitExampleCTAO} shows a comparison between the predicted flux of photons and the \textsc{Ctao} sensitivity at the GB (left panel) and the \textsc{Magic} observations of M15 (right panel). The impact of the different systematic uncertainties are also shown as colored bands in the plot.

\subsection{Upper limits on $\eta_e/\eta_\gamma$ for other $e^\pm$ injection scenarios}

\begin{figure}[t]
    \centering
    \includegraphics[width=0.49\linewidth]{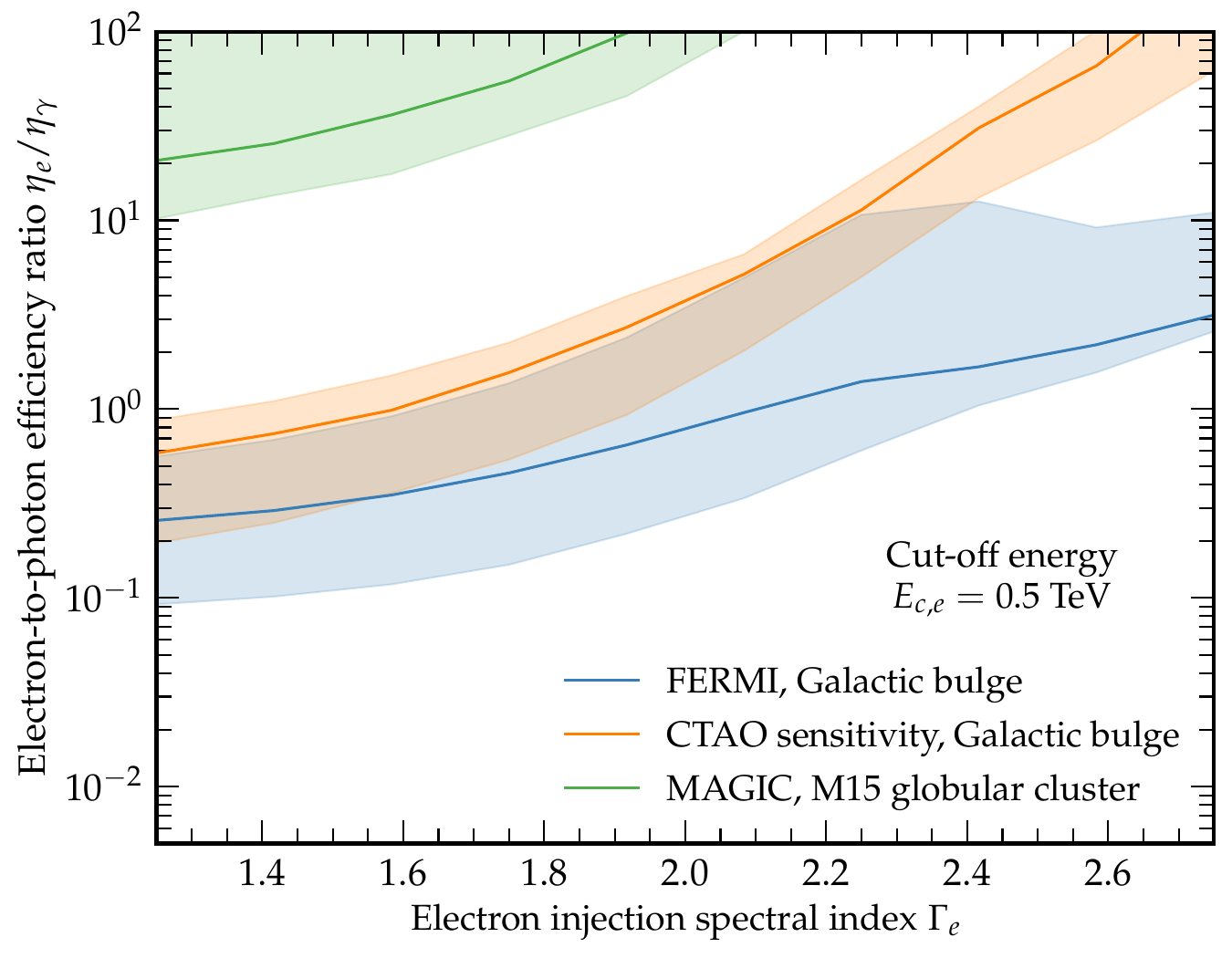}
    \includegraphics[width=0.49\linewidth]{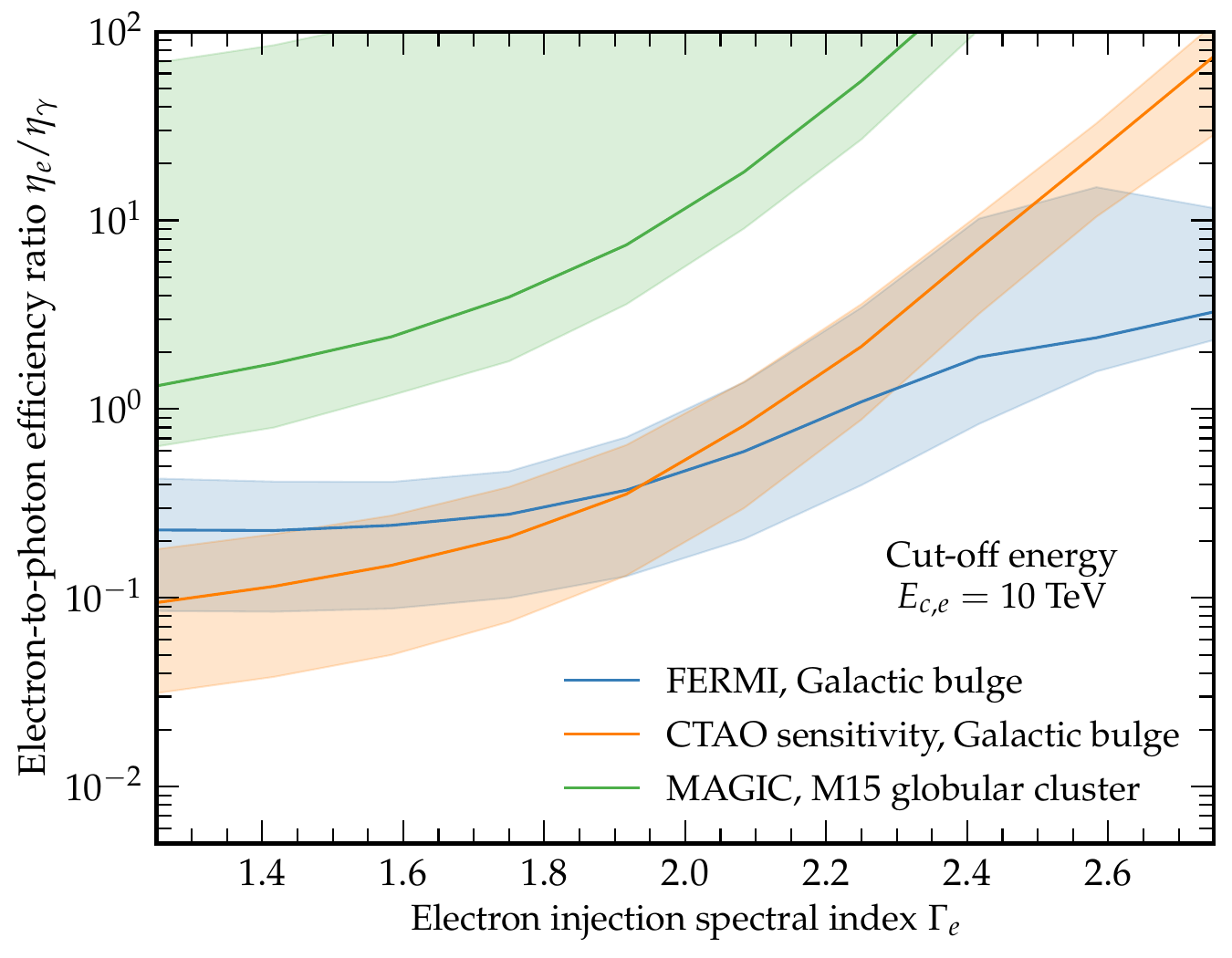}
    \includegraphics[width=0.49\linewidth]{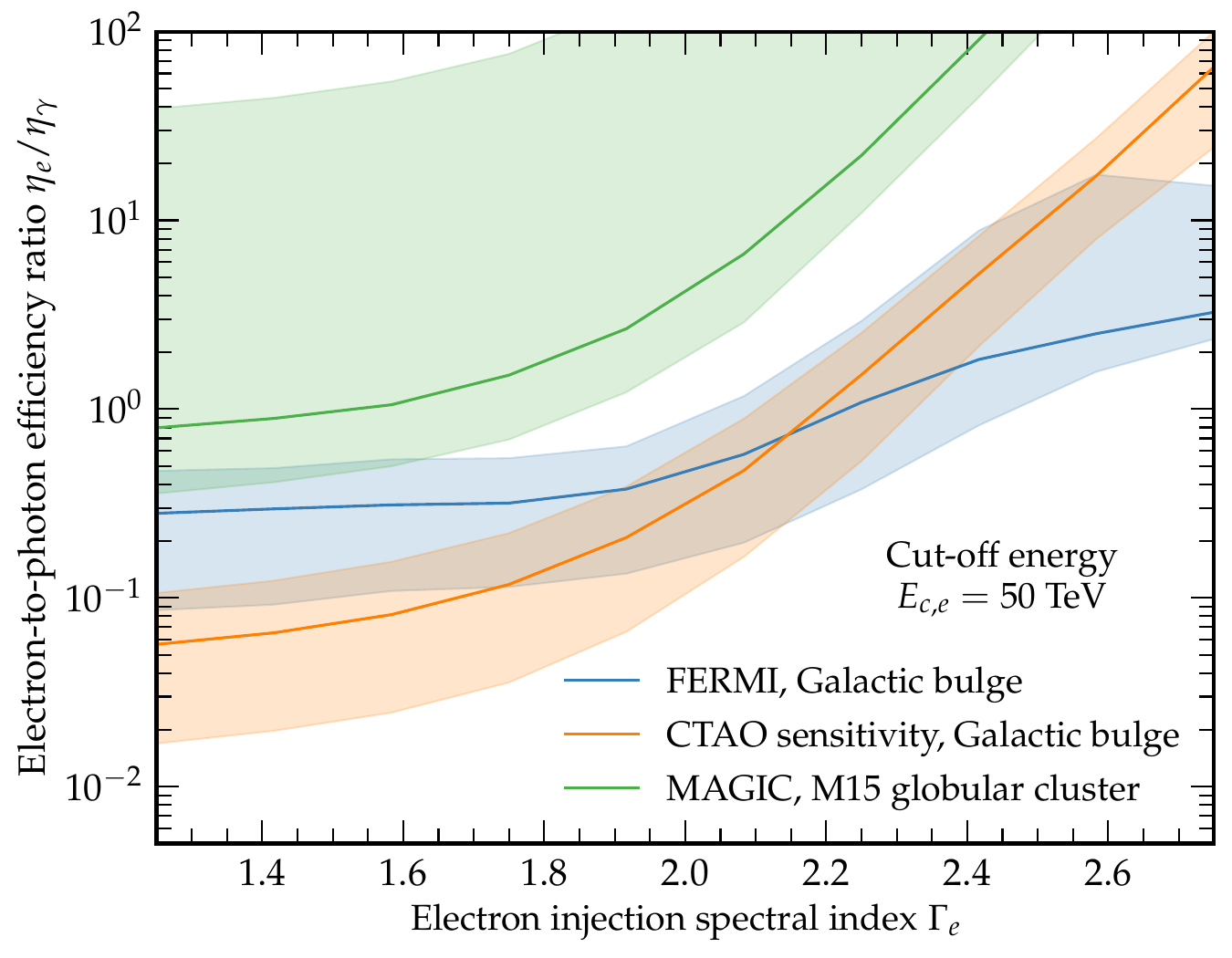}
    \includegraphics[width=0.49\linewidth]{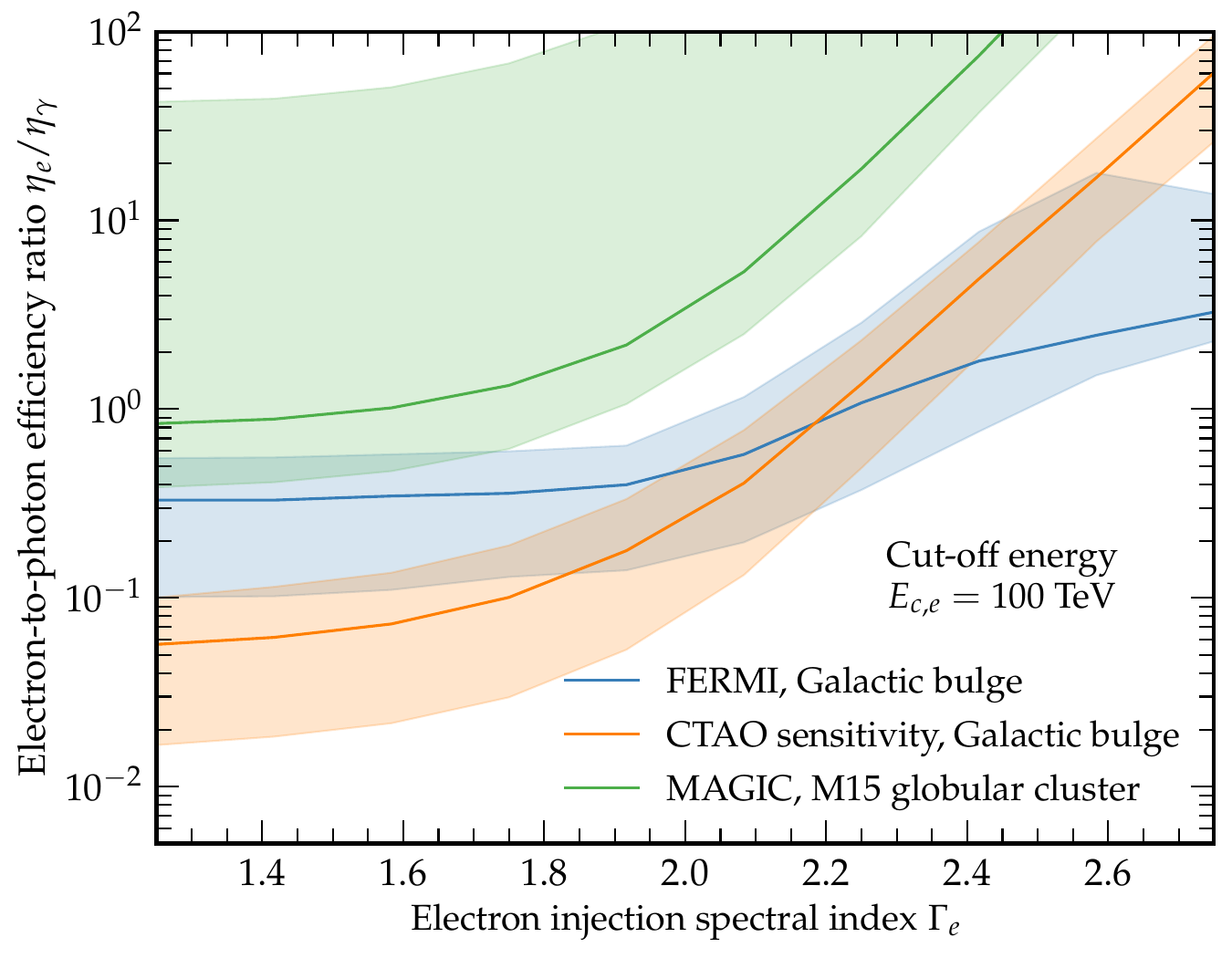}
    \caption{Same as Fig.~\ref{fig:moneyplot}, except when adopting different values of the $e^\pm$ injection cut-off energy $E_{c,e}$: 0.5 TeV (upper left panel), 10 TeV (upper right panel), 50 TeV (lower left panel), and 100 TeV (lower right panel).}
    \label{fig:FERMI_vs_CTAO}
\end{figure}

As a benchmark, we adopted a value of the $e^\pm$ injection cut-off energy $E_{c,e}$ to be 50 TeV. However, our conclusions remain the same when adopting other values of $E_{c,e}$, ranging from 0.5 to 100 TeV. This is illustrated by Fig.~\ref{fig:FERMI_vs_CTAO}.

\end{document}